\documentclass[11pt,english,twoside]{article}

\usepackage[T1]{fontenc}
\usepackage[latin1]{inputenc}
\usepackage[english]{babel}
\usepackage{lmodern}
\usepackage{a4wide}
\usepackage{amssymb, amsmath, amsthm}
\usepackage{slashed}
\usepackage{float}
\usepackage{graphicx}
\usepackage{psfrag}
\usepackage{lscape}
\usepackage[all]{xy}
\usepackage{hyperref}
\usepackage{enumerate}
\usepackage{dsfont}
\usepackage{cite}
\usepackage{mathabx}
\usepackage{color}
\usepackage{upgreek}

\newcommand{\beq}{\begin{equation}}
\newcommand{\eeq}{\end{equation}}
\def\bea#1\eea{\begin{align}#1\end{align}}
\newcommand{\nn}{\nonumber}
\newcommand{\w}{\wedge}
\newcommand{\id}{\mathds{1}}

\renewcommand{\i}{\ensuremath{\textnormal{i}}}

\def\del {\partial}
\def\d {{\rm d}}
\def\R {\mathcal{R}}
\def\L {\mathcal{L}}
\def\hhh {\mathcal{H}}
\def\Z {\mathbb{Z}}
\def\g {\gamma}
\def\tg {\widetilde{\gamma}}

\DeclareMathOperator{\re}{Re}
\DeclareMathOperator{\im}{Im}

\makeatletter
\@addtoreset{equation}{section}
\makeatother

\begin{document}

% ---------------------------------------- Title page
\begin{titlepage}

\rightline{\small LMU-ASC 82/12}
\rightline{\small MPP-2012-152}
\rightline{\small CERN-PH-TH/2012-235}

\vskip 2.1cm

{\fontsize{18.2}{21}\selectfont
  \flushleft{\noindent\textbf{(Non-)commutative closed string\\  [0.3cm]
  on T-dual toroidal backgrounds}} }

\vskip 0.2cm
\noindent\rule[1ex]{\textwidth}{1pt}
\vskip 0.9cm

\noindent\textbf{David Andriot$^{a,b}$, Magdalena Larfors$^{c}$, Dieter L\"ust$^{a,b,d}$, Peter Patalong$^{a,b}$}

\vskip 0.6cm
\begin{enumerate}[$^a$]
\item \textit{Arnold-Sommerfeld-Center for Theoretical Physics\\Fakult\"at f\"ur Physik, Ludwig-Maximilians-Universit\"at M\"unchen\\Theresienstra\ss e 37, 80333 M\"unchen, Germany}
\vskip 0.2cm
\item \textit{Max-Planck-Institut f\"ur Physik\\F\"ohringer Ring 6, 80805 M\"unchen, Germany}
\vskip 0.2cm
\item \textit{Mathematical Institute, Oxford University \\
24-29 St Giles', Oxford, OX1 3LB, England}
\vskip 0.2cm
\item \textit{CERN, Theory Group, 1211 Geneva 23, Switzerland}
\end{enumerate}
%
%\noindent Email:\\
\noindent {\small{\texttt{andriot@mpp.mpg.de, larfors@maths.ox.ac.uk,\\dieter.luest@lmu.de, peter.patalong@physik.uni-muenchen.de}}}

\vskip 2.1cm

\begin{center}
{\bf Abstract}
\end{center}

\noindent In this paper we investigate the connection between (non-)geometry and (non-)commutativity of the closed string. To this end, we solve the classical string on three T-dual toroidal backgrounds: a torus with $H$-flux, a twisted torus and a non-geometric background with $Q$-flux. In all three situations we work under the assumption of a dilute flux and consider quantities to linear order in the flux density. Furthermore, we perform the first steps of a canonical quantization for the twisted torus, to derive commutators of the string expansion modes. We use them as well as T-duality to determine, in the non-geometric background, a commutator of two string coordinates, which turns out to be non-vanishing. We relate this non-commutativity to the closed string boundary conditions, and the non-geometric $Q$-flux.

\vfill

\end{titlepage}

% ---------------------------------------- Table of contents
\tableofcontents
\newpage

\section{Introduction}
Strings are extended objects, and have a different perception of geometry than point particles do. The simple fact that a string can wind around a compact dimension gives string theories a wealth of interesting properties. Firstly, there is T-duality, which states that every string theory that is compactified on a torus has a physically equivalent description on a dual torus. Secondly, by applying T-duality to string compactifications with background fluxes, one can obtain non-geometric situations, where the internal part of space-time is no longer a standard manifold \cite{Hellerman:2002ax,Dabholkar:2002sy}. Finally, both open and closed string theories can become non-commutative in certain settings. In this paper, we study that assertion for closed strings propagating in a non-geometric background.

Phenomenological motivations for non-geometric string configurations were found in an analysis of four-dimensional supergravity (SUGRA) \cite{Shelton:2005cf} (see \cite{Andriot:2011uh} for a recent review on non-geometry and more references). Usually, such theories arise from compactifications of the ten-dimensional low-energy SUGRA description of string theory. If there are fluxes in the compact directions, a superpotential is generated in the four-dimensional theory.  Since such four-dimensional theories stem from string compactifications, they were expected to transform into each other under T-duality. However, such a duality could only be established if new terms were introduced in the superpotential.\footnote{Equivalently, the gauge algebra of the four-dimensional gauged SUGRA could only be made T-duality covariant if new structure constants were introduced \cite{Shelton:2005cf,Dabholkar:2005ve}.} The quantities generating these new terms were dubbed non-geometric $Q$- and $R$-fluxes, and should 
be T-duals to quantities in the Neveu--
Schwarz sector of the string. More precisely, the T-duality chain needed for a covariant four-dimensional superpotential is
\begin{equation}\label{eq:TdualityChain}
H_{abc} \stackrel{T_{a}}{\longrightarrow} f^{a}{}_{bc}
\stackrel{T_{b}}{\longrightarrow} Q_{c}{}^{ab}
\stackrel{T_{c}}{\longrightarrow} R^{abc}\, .
\end{equation}
In each step of this chain, a T-duality transformation is performed on direction $a,b$ and $c$, respectively. In this paper, our focus is on backgrounds with $Q$-flux, which has a local geometric description. The more exotic $R$-flux configurations are obtained when T-dualising on directions that are not isometries, and geometry should then be lost even locally \cite{Shelton:2005cf, Dabholkar:2005ve}.

While the $H$-flux and the structure constant $f$ have a clear ten-dimensional interpretation\footnote{$H$ is the exterior derivative of the Kalb--Ramond field $B$ and $f$ is related to the spin connection, and so indicates non-zero curvature. For this reason, $f$ is also known as the geometric flux.} the non-geometric fluxes did not at the time. The Neveu--Schwarz sector of ten-dimensional SUGRA  contains precisely two types of fluxes, $H$ and $f$, and so the origin of $Q$ and $R$ remained obscure. This was amended recently using a SUGRA field redefinition, which yields a globally defined ten-dimensional $Q$-flux in specific situations \cite{Andriot:2011uh}. Following this strategy, and also adopting the tools of double field theory \cite{DFT}, this result was extended and led to a ten-dimensional action and expressions for both $Q$ and $R$ \cite{Andriot:2012wx,Andriot:2012an}.\\

Independently of these studies of effective SUGRA descriptions of the string, related properties of string world-sheet theories have been investigated. In particular, the commutativity of the world-sheet fields, and the associated space-time geometry, in the presence of background fluxes has been under scrutiny. This led to the discovery of non-commutativity, when it was shown that the boundary theory of an open string ending on a D-brane with either constant $B$-field or an abelian gauge field is non-commuting \cite{open1,open2}.

In contrast, closed string theories are expected to remain commutative as long as the background is geometric, and more exotic set-ups seem to be required to find non-commutativity. Indeed, recently evidence has been found for a connection between non-geometry and closed string non-commutativity and even non-associativity \cite{Blumenhagen:2010hj,Lust:2010iy,Blumenhagen:2011ph,Condeescu:2012sp}. Investigations adopting a K-theory perspective in \cite{Bouwknegt:2000qt} also indicate that the closed string geometry becomes non-commutative on such backgrounds \cite{Saemann:2012ex}.\footnote{See \cite{Grange:2006es} for a different interpretation of these non-commutative theories.} Approaches using dual membrane theories \cite{Mylonas:2012pg} and matrix models \cite{Chatzistavrakidis:2012qj} arrive at the same conclusion.

It was shown in \cite{Lust:2010iy,Condeescu:2012sp}, for a non-geometric background with elliptic monodromy (which can be viewed as a freely acting asymmetric orbifold), that a non-vanishing commutator of closed string coordinates is proportional to the non-geometric flux times a winding number. Guided by these results, we formalised in \cite{Andriot:2012an} the connection between non-geometric fluxes and closed string non-commutativity through the conjecture
\begin{equation}\label{commcloseda}
 [ {\cal X}^{\mu}, {\cal X}^{\nu} ]_{\rm closed} \sim  \oint_{C_{\rho}}Q_{\rho}{}^{\mu \nu}({\cal X})~ \d{\cal X}^{\rho} \ .
\end{equation}
The scope of this paper is to show that this conjecture holds in one of the most famous examples of non-geometric set-ups: the $Q$-flux dual of the flat three-torus with $H$-flux \cite{Kachru:2002sk,Lowe:2003qy}. This configuration, and its T-duals, have been studied at length in the literature (see \cite{Andriot:2011uh,Blumenhagen:2011ph} for recent accounts with focus on non-geometry and non-associativity). We will recapitulate some salient features here, and also clarify how, by assuming a sufficiently dilute flux, these set-ups can be approximations of string backgrounds.

The flat three-torus with $H$-flux has two isometries: the Kalb--Ramond field $B$ necessarily depends on one of the torus coordinates in order to give a non-trivial $H$-flux. It is useful to describe the configuration as a (trivial) fibration of a two-torus over a base circle, and let $B$ live along the fibre and depend on the base circle coordinate.  T-dualising on one of the fibre directions, one finds (using the Buscher rules \cite{Buscher:1987sk,Buscher:1987qj}) a twisted torus with zero $H$-flux, whose fibre twisting is related to a non-trivial geometric flux $f$. T-dualising on the second fibre direction, one reaches a non-geometric situation where the background metric and $B$-field are not globally well-defined. Indeed, T-duality is required to act as transition functions for these fields, and the configuration can be viewed as a T-fold \cite{Hull:2004in}.  As shown in \cite{Andriot:2011uh}, a description with a globally defined metric and $Q$-flux exists for this set-up. Thus, this toroidal example 
perfectly matches the T-duality chain 
\eqref{eq:TdualityChain} (and was, in fact, an inspiration for it).

To study strings on this field configuration, it should better be a consistent string background, and so in particular Weyl invariant. Thus, the one-loop $\beta$-functions of the string sigma-model should vanish, which corresponds to satisfying the SUGRA equations of motion \cite{Green:1987sp}. For the flat torus with $H$-flux, this is generically not the case. In particular, the dilaton and Einstein equations restricted to the Neveu--Schwarz sector are given by
\bea
& \R  + 4 (\nabla^2 \phi - (\del \phi)^2) = \frac{1}{12} H_{\mu\nu\rho} H^{\mu\nu\rho}  \label{eq:dileom}\\
& \R_{\mu\nu} -\frac{G_{\mu\nu}}{2} \R + 2 \nabla_{\mu}\del_{\nu} \phi - 2 G_{\mu\nu} (\nabla^2 \phi - (\del \phi)^2) = \frac{1}{4} \left(H_{\mu\kappa\lambda}  {H_{\nu}}^{\kappa\lambda}
-\frac{G_{\mu\nu}}{6} H_{\rho\kappa\lambda} H^{\rho\kappa\lambda}  \right)  \ . \label{eq:Einstein}
\eea
For a flat torus with constant dilaton but non-zero $H$, these equations are violated due to terms proportional to the square of the $H$-flux. One way to make sense of this configuration is to complete it with more ingredients (Ramond-Ramond fluxes, orientifold planes) in the other dimensions not considered here; this has been done in \cite{Marchesano:2007vw} to get solutions of type IIB SUGRA as those of \cite{Giddings:2001yu}. We will proceed in a different manner. We only consider NSNS sector contributions to the SUGRA equations of motion, and assume the torus volume is so large that
\beq
\frac{H}{R_1 R_2 R_3} \ll 1 \; .\label{eq:approx}
\eeq
Here, $R_{\nu=1,2,3}$ denote the radii of the torus, and $H$ the flux component $H_{123}$ (our conventions are explained below \eqref{action} and above \eqref{Ametric}). As we show in detail in section \ref{sec:2}, the quadratic flux terms in \eqref{eq:dileom} and \eqref{eq:Einstein} are quadratic in this small flux density, and so can be neglected to a good approximation. Thus, using this ``dilute flux approximation'',\footnote{In the twisted torus, the approximation could also be called a weak curvature approximation. Approximations of this type have been used before, see e.g. \cite{Blumenhagen:2011ph, Davidovic:2012de} and references therein for related discussions.} this field configuration can be viewed as an approximate string background. The result can easily be extended to the two T-dual set-ups (T-duality actually guarantees that if it is true for one, it remains true for the others).\footnote{The asymmetric orbifold studied in \cite{Condeescu:2012sp} is an example of an exact CFT solution where non-commutativity appeared to be related to the underlying non-geometry. Thanks to the approximation considered here, we have at hand another example with similar characteristics (a non-geometric set-up); it should therefore be enough to recover the non-commutativity features, and we leave the study of a background beyond this approximation to future investigations.}

Consequently, for a dilute flux, we find that the torus with $H$-flux and the twisted torus are both approximate geometric string backgrounds. It therefore makes sense to analyse the properties of their world-sheet theories in detail, and we will do so in this paper. We will derive the equations of motion and boundary conditions for the world-sheet fields and find their solutions to linear order in the flux density. We will also investigate the T-duality relations between equations of motions, coordinate solutions and canonical commutators between coordinates. This is presented in section \ref{sec:2}, and shows that T-duality is a remarkable tool: not only does it relate the classical coordinate solutions of the two geometric backgrounds, but it also maps their quantum properties, i.e. most of their operator commutators.

However, a similar direct analysis of the non-geometric $Q$-flux background is not possible. This is especially pertinent for the commutators of the coordinates. While it is natural to impose canonical commutator relations in any geometric frame, the same does not hold in a non-geometric setting. Thus, in these settings, the commutators of coordinates are a priori undetermined. To compute them, we hence proceed indirectly through T-duality, using that both the classical coordinate solutions and their commutator relations are T-dual to well-understood geometric quantities. Thus, by T-dualising the twisted torus solution $Y^{\mu}(\tau,\sigma)$, we obtain the coordinate solutions $Z^{\mu}(\tau,\sigma)$ of the non-geometric situation as a (complicated) combination of zero modes and oscillator modes. The commutation relations of these modes are obtained from a (partial) canonical quantization of the twisted torus solution. As a result, we find the following non-vanishing commutator
\beq
[Z^1 (\tau,\sigma) , Z^2 (\tau,\sigma')] \xrightarrow{\sigma' \rightarrow \sigma}  - \frac{\i}{2} \frac{\pi^2}{3} N^3 H \ .
\eeq

As expected, our result shows that the non-commutativity is contingent on the flux and winding. The explicit expression for the $Q$-flux, that was derived in \cite{Andriot:2011uh}, shows that $Q$ is proportional to $H$, and so the analysis confirms the conjecture \eqref{commcloseda}. In particular, if $H$ is zero, we have a geometric and commutative situation. Furthermore, just as suggested by the integral in the conjecture, it is the extension of the string that is the source of non-commutativity. Indeed, if we put the winding number of our solution to zero, we loose non-commutativity. The relation to the extension of the string is not surprising, and shows that the non-commutativity is a non-local effect. 

The rest of this paper is organised as follows. In section \ref{sec:2} we perform a thorough analysis of the T-dual configurations. We show that they are approximate string backgrounds when the background flux density is small, and study the T-duality relations between equations of motion and canonical commutators. In section \ref{sec:geo} we solve the equations of motion in the two geometric situations to linear order in the flux density, and discuss the T-duality relations among their solutions. We then perform a partial canonical quantization for the twisted torus background, deriving the commutation relations of the modes of its solution.  In section \ref{sec:nongeo}, we study the non-geometric background. Using T-duality we find its coordinate solutions, and work out their commutators, leaving some details to appendix \ref{sec:fixcom}. We study the origin of the non-commutativity and compare with the geometric backgrounds. We also relate the non-commutativity to the $Q$-flux. Finally, we summarize our 
conclusions and give suggestions for future investigations. Appendices \ref{ap:not} and \ref{ap:Td} explain our notation and target space aspects of T-duality. In appendix \ref{sec:monod}, we comment on the relations between monodromies, closed string boundary conditions and non-geometry.

Since this paper is on the long side, let us decompose it into three themes. A reader interested in non-commutativity should focus on sections \ref{sec:ZZ} and \ref{sec:ZZdisc} as well as appendix \ref{sec:monod}, while one more interested in the canonical quantization of the twisted torus should look at \ref{sec:classstring}, \ref{sec:quantgen} and \ref{sec:geo}. Finally, \ref{subsubsec:Htorus}, \ref{subsubsec:Ttorus}, \ref{sec:Tdual} and \ref{sec:solutions} discusses the classical string on the two geometric backgrounds.

\section{Classical and quantum string on the T-dual backgrounds}\label{sec:2}

In this section, our starting point is the standard world-sheet action
\begin{equation}
S=-\frac{1}{4\pi\alpha'}\int_\Sigma \d^2\sigma \left( G_{\mu \nu}({\cal X})\ \eta^{\alpha\beta} + B_{\mu \nu}({\cal X})\ \varepsilon^{\alpha\beta} \right) \partial_\alpha {\cal X}^{\mu} \partial_\beta {\cal X}^{\nu} \ , \label{action}
\end{equation}
where the metric $G$ and the $B$-field specify which background we consider. The world-sheet metric $\eta_{\alpha \beta}$ is Minkowski with signature $\eta_{\tau \tau}=-\eta_{\sigma\sigma}=-1$, and we take as a convention $\varepsilon_{\tau \sigma}= -\varepsilon_{\sigma \tau} = 1$. In addition, we fix for convenience $\alpha'=\frac{1}{2}$.

From this action, we first recall Buscher's T-duality procedure \cite{Buscher:1987sk,Buscher:1987qj}, and deduce the T-duality relations between string coordinates. We then present the three T-dual toroidal backgrounds, and study the associated world-sheet equations of motion and closed string boundary conditions.\footnote{In appendix \ref{sec:monod}, the global properties of these backgrounds are rephrased in terms of monodromies, out of which we propose a way to derive the boundary conditions of (doubled) string coordinates.} Finally, we come back to the T-duality relations among coordinates of the different backgrounds and study how they relate the equations of motions and boundary conditions, as well as the canonical commutation relations that we first introduce.

\subsection{Buscher's approach to T-duality}

Let us first briefly sketch the procedure introduced by Buscher \cite{Buscher:1987sk,Buscher:1987qj}, in order to derive the T-duality transformation rules we need. For more details, see for instance \cite{Thompson:2010sr}. We start with the initial world-sheet action \eqref{action}, that can be rewritten as follows
\beq
S= \frac{2}{\pi} \int_{\Sigma} \d^2 \sigma \ E_{\mu \nu}({\cal X}) \ \del_{\sigma_-} {\cal X}^{\mu}\ \del_{\sigma_+} {\cal X}^{\nu} \ , \ {\textrm{with}}\ E_{\mu \nu}= G_{\mu \nu} + B_{\mu \nu} \ ,
\eeq
where we used the conventions given below \eqref{action}, introduced $\sigma_{\pm}=\tau \pm \sigma$ and $2 \del_{\sigma_{\pm}}=\del_{\tau} \pm \del_{\sigma}$.

To perform a T-duality along $\mu=\iota$, one needs this direction to be an isometry, meaning in practice that background fields $E_{\mu \nu}({\cal X})$ do not depend on ${\cal X}^{\iota}$ (we denote by $\mu=\kappa,\lambda$ the other directions). The first step of the procedure then consists in gauging this isometry, by adding to $\del_{\sigma_{\pm}} {\cal X}^{\iota}$ gauge fields $A_{\pm}$, and to the initial action $S$ the following piece
\beq
S_A= -\frac{2}{\pi} \int_{\Sigma} \d^2 \sigma \ \hat{{\cal X}}\ \left(\del_{\sigma_-} A_+ - \del_{\sigma_+}  A_- \right) \ .
\eeq
Here, the field $\hat{{\cal X}}$ is a Lagrange multiplier, which will become the T-dual coordinate. The fields $A_{\pm}$ can be thought of as light-cone gauge fields; it is thus their field strength which appears here. The equation of motion for $\hat{{\cal X}}$ puts this field strength to vanish, so that these gauge fields are pure gauge. Imposing this constraint, and shifting ${\cal X}^{\iota}$ (possible thanks to the isometry), one recovers the initial theory given by the action $S$.

One can equivalently integrate out the gauge fields, instead of $\hat{{\cal X}}$. To do so, one should first derive their equations of motion. Note that these are the same as the equations for $\del_{\sigma_{\pm}} {\cal X}^{\iota} + A_{\pm}$ (in particular one can replace $A_{\pm}$ by this sum within $S_A$, without changing it). To integrate out $A_{\pm}$, one then in practice replaces in the full action these quantities by their on-shell value. By construction, the resulting theory is equivalent to the starting one: we say that the new action is T-dual to $S$. Up to a total derivative, this new action has the same form as $S$, where one replaces ${\cal X}^{\iota}$ by $\hat{{\cal X}}$, and the field $E_{\mu \nu}$ by $\hat{E}_{\mu \nu}$, given as follows
\beq
\hat{G}_{\iota \iota}=\frac{1}{G_{\iota \iota}}\, ,\quad
\hat{E}_{\iota \kappa}=-\frac{E_{\iota \kappa}}{G_{\iota \iota}}\, ,\quad
\hat{E}_{\kappa \iota}=\frac{E_{\kappa \iota}}{G_{\iota \iota}}\, ,\quad
\hat{E}_{\kappa \lambda}=E_{\kappa \lambda}-\frac{E_{\kappa \iota} E_{\iota \lambda}}{G_{\iota \iota}}\, .\label{BuscherE}
\eeq
These are the well-known Buscher rules \cite{Buscher:1987sk,Buscher:1987qj} that describe how the target space fields transform under T-duality. We rediscuss these target space transformations in appendix \ref{ap:Td}.

Let us come back to the equations of motion derived for $A_{\pm}$ (equivalently for $\del_{\sigma_{\pm}} {\cal X}^{\iota}+A_{\pm}$). These equations are the core of the procedure and allow to relate the T-dual situations. As before, performing a trivial shift on ${\cal X}^{\iota}$ would absorb a pure gauge $A_{\pm}$ into $\del_{\sigma_{\pm}} {\cal X}^{\iota}$; equivalently one can choose to gauge-fix $A_{\pm}$ to zero. In both cases, the equations of motion simplify to\footnote{On the matter of gauge-fixing, see \cite{Nibbelink:2012jb}. On the relations between T-dual coordinates, see also \cite{Duff:1989tf}.}
\bea
& \del_{\sigma_+} \hat{{\cal X}} =\ \ G_{\iota \iota}\ \del_{\sigma_+} {\cal X}^{\iota} + E_{\iota \kappa}\ \del_{\sigma_+} {\cal X}^{\kappa}\label{Aeom+} \\
& \del_{\sigma_-} \hat{{\cal X}} =- G_{\iota \iota}\ \del_{\sigma_-} {\cal X}^{\iota} - E_{\kappa \iota }\ \del_{\sigma_-} {\cal X}^{\kappa}\ .\label{Aeom-}
\eea
Consequently, one deduces the following relations between derivatives of T-dual coordinates\footnote{Note that in the whole procedure, another set of conventions is possible. In particular, one can take the opposite sign in front of $S_A$, and the opposite sign for $\varepsilon_{\tau \sigma}$. This results in having the opposite sign for the off-diagonal pieces in \eqref{BuscherE}, and in exchanging the indices of these same pieces in \eqref{Aeom+} and \eqref{Aeom-}. One can of course choose such conventions, but they will be incompatible with the conventions of \eqref{BuscherH}.}
\bea
\partial_{\tau} \hat{{\cal X}}^{\iota}&=G_{\iota \iota}\partial_{\sigma} {\cal X}^{\iota}+G_{\iota \kappa}\partial_{\sigma} {\cal X}^{\kappa}+B_{\iota \kappa}\partial_{\tau} {\cal X}^{\kappa} \label{Tdcoordrel}\\
\partial_{\sigma} \hat{{\cal X}}^{\iota}&=G_{\iota \iota}\partial_{\tau} {\cal X}^{\iota}+G_{\iota \kappa}\partial_{\tau} {\cal X}^{\kappa} +B_{\iota \kappa}\partial_{\sigma} {\cal X}^{\kappa}\nn \\
\partial_{\tau} \hat{{\cal X}}^{\kappa}&=\partial_{\tau}  {\cal X}^{\kappa}\, ,\quad\partial_{\sigma} \hat{{\cal X}}^{\kappa}=\partial_{\sigma}{\cal X}^{\kappa} \ .\nn
\eea
We will make use of these relations, but let us first present the different T-dual backgrounds we will work with.

\subsection{The classical string on the different T-dual backgrounds}\label{sec:classstring}

In this section, we present the different T-dual backgrounds, and derive for each of them the world-sheet equations of motion. We also discuss the boundary conditions of the coordinates.

\subsubsection{Torus with $H$-flux}\label{subsubsec:Htorus}

We first consider three dimensions of the target space to be a flat torus along ${\cal X}^{\mu}=X^{\mu=1,2,3}$ with periodic identifications $X^{\mu} \sim X^{\mu} + 2 \pi$, and a $H$-flux $H_3=H \d X^1 \w \d X^2 \w \d X^3$, where $H$ is a constant. We only consider the target space fields along the torus; the metric is then given in terms of the radii $R_{\mu}$ by
\begin{equation} \label{Ametric}
G = \begin{pmatrix} R_1^2 & 0 & 0 \\ 0 & R_2^2 & 0 \\ 0 & 0 & R_3^2 \end{pmatrix} \ ,
\end{equation}
and the $B$-field is fixed to a particular gauge that is linear in $X^3$,
\begin{equation} \label{Abfield}
B_{12} = - B_{21} = H X^3 \ , \quad B_{13} = B_{31} = B_{23} = B_{32} = 0 \ .
\end{equation}

As discussed in the introduction, for this configuration of fields to be a valid string theory background, the SUGRA equations of motion should be satisfied. With a constant dilaton, the equations \eqref{eq:dileom} and \eqref{eq:Einstein} reduce here to
\beq
\left(\frac{H}{R_1 R_2 R_3}\right)^2 = 0 \ ,
\eeq
where we used that in curved indices, $H_{123}=H$. Thus, these equations of motion are satisfied up to linear order in $H/R_1 R_2 R_3$. This can be realised physically by considering a sufficiently large torus which would dilute the flux, and assure that \eqref{eq:approx} is satisfied. Note that the flux, i.e. $H$, is quantized in string theory. The flux density $H/R_1 R_2 R_3$ can be small, and we will work with this approximation of the dilute flux throughout the paper, i.e. at linear order in the previous quantity. Finally, we should also consider the $B$-field equation of motion and the $H$-flux Bianchi identity, given respectively by the forms
\beq
\d (e^{-2\phi} * H_3) = 0 \ , \ \d H_3  =0 \ .
\eeq
While the latter is trivially satisfied for our constant $H$-flux, the former is less straightforward, because the Hodge star $*$ involves the volume and the orthogonal metric components. Here we restrict ourselves to the three-dimensional flat torus with a constant dilaton, and the equation is satisfied because all fields are constant. In a more general case, other assumptions will have to be made.

Given this background, we now study the closed string living on it. Starting from \eqref{action} and its associated conventions, using \eqref{Ametric} and \eqref{Abfield}, we derive the equation of motion for a closed string moving on a torus with $H$-flux:
\beq
\partial_{\alpha} \partial^{\alpha} X^{\mu} (\tau,\sigma) = G^{\mu \lambda} H_{\lambda \nu \rho} \partial_{\sigma} X^{\nu} \partial_{\tau} X^{\rho} \; .
\eeq
In order to simplify notations, and the approximation, we rescale all quantities (see table \ref{tab:resc} in appendix \ref{ap:not}), which results in
\beq
X^{\mu} \to \frac{1}{R_{\mu}} X^{\mu} \; \mbox{ (no sum)}\ , \ G_{\mu \nu} \to \eta_{\mu \nu} \ , \
H \to H R_1 R_2 R_3 \; .
\eeq
These rescaled quantities will be used in the remainder of the paper. We then have the rescaled equations of motion
\beq \label{Aeom}
\partial_{\alpha} \partial^{\alpha} X^{\mu} (\tau,\sigma) =  H \epsilon^{\mu}{}_{\nu \rho} \partial_{\sigma} X^{\nu} \partial_{\tau} X^{\rho} \; ,
\eeq
where $\epsilon^{\mu}{}_{\nu \rho} = \eta^{\mu \lambda}\epsilon_{\lambda \nu \rho}$. In this rescaled notation, the approximation \eqref{eq:approx} translates simply into retaining terms up to linear order in $H$, which is now to be regarded as a flux density: $(H)_{\rm new} = (H)_{\rm old}/R_1R_2R_3$. Note as well that this new $H$ is equal to $H_{123}$ in flat indices, i.e. the one that enters the T-duality chain \eqref{eq:TdualityChain}. This is a first hint that this quantity will appear and play the same role in the other T-dual backgrounds.

Finally, let us consider the following boundary conditions
\beq \label{eq:bdyX}
X^{\mu}(\tau,\sigma+2\pi) = X^{\mu}(\tau,\sigma) + 2 \pi N_X^\mu \; ,
\eeq
that are compatible with the torus periodic identifications. These boundary conditions will be completed in \eqref{eq:bdyXmonod} with those of dual coordinates, using a doubled formalism.

We index here the winding modes $N_X^{\mu}$ by $X$ to distinguish them from the modes $N^{\mu}$ in the twisted torus configuration. The same distinction will be made on the other modes; in particular, to ease the notations, no specific index will be carried by the modes of the twisted torus, as those are the ones mostly used in the paper. Let us now turn to this second configuration.

\subsubsection{Twisted torus}\label{subsubsec:Ttorus}
Secondly, we consider a twisted torus along ${\cal X}^{\mu}=Y^{\mu=1,2,3}$ with metric
\begin{equation}
G  = \begin{pmatrix} \frac{1}{R_1^2} & - \frac{HY^3}{R_1^2} & 0 \\ - \frac{HY^3}{R_1^2} & R_2^2 + \left(\frac{HY^3}{R_1}\right)^2 & 0 \\ 0 & 0 & R_3^2 \end{pmatrix} \ .\label{metric}
\end{equation}
This is a three-dimensional nilmanifold generated by the Heisenberg algebra, and its Maurer-Cartan one-forms, in particular $(\d Y^1 - HY^3 \d Y^2)/R_1$, are globally well-defined provided the following identifications are satisfied\footnote{See \cite{Andriot:2010ju} for a review on nil- and solvmanifolds, their compactness and associated discrete identifications.}
\beq
(Y^1, Y^2, Y^3) \sim (Y^1 + 2 \pi, Y^2, Y^3) \sim (Y^1, Y^2 + 2 \pi, Y^3) \sim (Y^1 + 2\pi H Y^2, Y^2, Y^3 + 2 \pi) \ . \label{monod1}
\eeq
In addition, we consider no $B$-field. Using \eqref{BuscherE}, or appendix \ref{ap:Td}, one can verify that this field configuration is related to the flat torus with $H$-flux by a T-duality in the ${\cal X}^1$-direction. This is true provided we identify $X^3=Y^3$, while the $H$ parameter and the radii we have here are the same as in \eqref{Ametric} and \eqref{Abfield}. We will return to this T-duality in section \ref{sec:Tdual}.

To identify the dilute flux approximation in this background, we again study the target space equations of motion. For a constant dilaton and no $H$-flux, equations \eqref{eq:dileom} and \eqref{eq:Einstein} become
\beq
\R =0 \ , \ \R_{\mu\nu} - \frac{1}{2} G_{\mu \nu} \R =0 \ .
\eeq
Using that the non-zero components of the Ricci tensor (in curved indices) and the Ricci scalar for the twisted torus are
\beq
\R_{11}=\frac{1}{2 R_1^2} \left(\frac{H}{R_1R_2R_3} \right)^2 \ , \ \R_{22 / 33}= - \frac{R_{2/3}^2}{2} \left(\frac{H}{R_1R_2R_3} \right)^2 \ , \ \R=- \frac{1}{2} \left(\frac{H}{R_1R_2R_3} \right)^2 \ ,
\eeq
we find, just as in the T-dual situation discussed above, that these equations are satisfied at linear order in $H/R_1R_2R_3$. Another way to see that the approximation remains \eqref{eq:approx}, is to notice that the (non-trivial) structure constant ${f^1}_{23}=-H/R_1R_2R_3$. As already mentioned for the flat torus, this is the quantity that appears in the T-duality chain \eqref{eq:TdualityChain}, and is considered small in our approximation.

It is straight forward to derive the world-sheet equations of motion in the twisted torus background. In order to simplify their appearance, and the approximation, we again rescale according to table \ref{tab:resc} in appendix \ref{ap:not}, giving in particular
 \beq
Y^1 \rightarrow R_1 Y^1 \ , \ Y^{2,3} \rightarrow \frac{1}{R_{2,3}} Y^{2,3} \ , \ H \rightarrow H R_1R_2R_3 \ . \label{scaling}
\eeq
Then, our approximation is again simply given by the linear order in $H$, while the rescaled equations of motion are
\bea
\partial_{\alpha} \partial^{\alpha} Y^1 &= H \left(
Y^3 \partial_{\alpha} \partial^{\alpha} Y^2 + \partial_{\alpha} Y^2 \partial^{\alpha} Y^3 \right) \label{eqY1}\\
\partial_\alpha \partial^\alpha Y^2 &= H  \left( \partial_\alpha Y^1 \partial^\alpha Y^3  -  H Y^3
\partial_\alpha Y^2 \partial^\alpha Y^3 \right) \label{eqY2} \\
\partial_{\alpha} \partial^{\alpha} Y^3 &= H \left(
- \partial_{\alpha} Y^1 \partial^{\alpha} Y^2 +
H Y^3 \partial_{\alpha} Y^2 \partial^{\alpha} Y^2 \right) \ . \label{eqY3}
\eea
They simplify, at linear order in $H$, to
\begin{align}
\partial_\alpha\partial^\alpha Y^{\mu} &= H \theta^{\mu}{}_{\nu\rho}\partial_\alpha Y^{\nu} \partial^\alpha Y^{\rho}  \ ,\label{eq:eomY}
\end{align}
where we introduce ${\theta^1}_{23}={\theta^2}_{13}=-{\theta^3}_{12}=1$, with all other components being zero.

The twisting of the torus, described by the identifications \eqref{monod1}, makes the boundary conditions for the coordinate fields non-trivial. Allowing for some winding $N^{\mu}$, we have
\begin{align}
Y^1(\tau,\sigma+2\pi) &= Y^1(\tau,\sigma) + 2\pi N^1 + 2\pi N^3 H Y^2 (\tau,\sigma)\label{bdy1} \\
Y^2(\tau,\sigma+2\pi) &= Y^2(\tau,\sigma) + 2\pi N^2 \label{bdy2} \\
Y^3(\tau,\sigma+2\pi) &= Y^3(\tau,\sigma) + 2\pi N^3 \ . \label{bdy3}
\end{align}
These boundary conditions will be completed in \eqref{eq:bdyYmonod} with those of dual coordinates, using a doubled formalism.

\subsubsection{The non-geometric background}\label{sec:nongeobg}

Finally, we consider a third situation with the string coordinates denoted ${\cal X}^{\mu}=Z^{\mu=1,2,3}$ and the following field configuration
\begin{equation}
G = f \begin{pmatrix}  \frac{1}{R_1^2} & 0 & 0 \\ 0 & \frac{1}{R_2^2} & 0 \\ 0 & 0 & \frac{R_3^2}{f} \end{pmatrix}\ , \ B = f \begin{pmatrix} 0 & -\frac{HZ^3}{R_1^2 R_2^2} & 0 \\ \frac{HZ^3}{R_1^2 R_2^2} & 0 & 0 \\ 0 & 0 & 0 \end{pmatrix}\ , \ f=\left(1+\left(\frac{H Z^3}{R_1R_2} \right)^2 \right)^{-1} \ .\label{eq:nongeofields}
\end{equation}
This configuration is known to be non-geometric in the sense of \cite{Hellerman:2002ax, Dabholkar:2002sy}. Indeed, when going around the circle along $Z^3$, one cannot find a diffeomorphism or a gauge transformation which would make these fields globally defined; on the contrary one can achieve this by using a T-duality as the transition function between two patches on the $Z^3$ circle \cite{Kachru:2002sk, Lowe:2003qy}.

This field configuration is also T-dual to the twisted torus. By performing a T-duality along ${\cal X}^2$, and identifying $Y^3=Z^3$, one obtains precisely the above fields using \eqref{BuscherE} or appendix \ref{ap:Td} (the parameter $H$ and the radii are the same as before). This implies that if the torus with $H$-flux or the twisted torus are consistent string backgrounds, or can be completed to such, the same should hold here. We infer that we can make the same approximation as \eqref{eq:approx}, so that the non-geometric field configuration becomes a background. There is an easy way to verify this: after rescaling quantities as indicated in table \ref{tab:resc} in appendix \ref{ap:not}, and considering at most the linear order in $H$, the non-geometric fields \eqref{eq:nongeofields} reduce to precisely those of the torus with $H$-flux, up to a change of sign of the $B$-field (see table \ref{tab:rescandH}), so they clearly satisfy the SUGRA equations.

Despite the similarity with this previous geometric background, we know that the fields here should only be considered locally, and that their global properties are non-trivial. The global aspects differ from those of the torus with $H$-flux, as we discuss in appendix \ref{sec:monod}, and similarly, the T-duality relations \eqref{eq:TdcoordXZ} among coordinates will indicate boundary conditions for the $Z^{\mu}$ that are different from those of the $X^{\mu}$. More generally, the idea of this paper is to make an intensive use of the T-duality relations between these various backgrounds to study the string properties on the non-geometric one.

\subsubsection*{Summary of the rescaled and approximated background fields}

As argued for each of the three T-dual backgrounds, the supergravity equations of motion are satisfied at linear order in the $H$ parameter. In table \ref{tab:rescandH}, we summarize all target space fields, namely \eqref{Ametric} and \eqref{Abfield}, \eqref{metric}, and \eqref{eq:nongeofields}, rescaled according to table \ref{tab:resc} in appendix \ref{ap:not}, and expanded at linear order in $H$.
\begin{table}[H]
\begin{center}
\begin{tabular}{|c||c|c|}
\hline
Backgrounds & ${\cal X}^{\mu}$ & Target space fields \\
\hline
\hline &&\\[-0.5ex]
Torus + $H$-flux & $X^{\mu}$ & $G = \begin{pmatrix} 1 & 0 & 0 \\ 0 & 1 & 0 \\ 0 & 0 & 1 \end{pmatrix}\ , \ B = \begin{pmatrix} 0 & H X^3 & 0 \\ - H X^3 & 0 & 0 \\ 0 & 0 & 0 \end{pmatrix} $ \\[5ex]
\hline  &&\\[-0.5ex]
Tw. torus & $Y^{\mu}$ & $G = \begin{pmatrix} 1 & - HY^3 & 0 \\ - HY^3 & 1 & 0 \\ 0 & 0 & 1 \end{pmatrix} + {\cal O}(H^2)\ , \ B = 0 $ \\[5ex]
\hline &&\\[-0.5ex]
Non-geom. & $Z^{\mu}$ & $G = \begin{pmatrix}  1 & 0 & 0 \\ 0 & 1 & 0 \\ 0 & 0 & 1 \end{pmatrix} + {\cal O}(H^2)\ , \ B =  \begin{pmatrix} 0 & -HZ^3 & 0 \\ HZ^3 & 0 & 0 \\ 0 & 0 & 0 \end{pmatrix} + {\cal O}(H^2)$  \\[5ex]
\hline
\end{tabular}
\caption{Rescaled background fields, keeping terms to linear order in $H$.}\label{tab:rescandH}
\end{center}
\end{table}
\vspace{-0.2in}
We recall that these fields are T-dual according to \eqref{BuscherE} or appendix \ref{ap:Td}, provided one identifies $X^3=Y^3=Z^3$ (on that point, see the discussion after \eqref{eq:TdcoordXZ}). These fields will be used in the remainder of the paper, in particular to compute the canonical momentum \eqref{canmom}.

\subsection{Relating the classical string on the different T-dual backgrounds}\label{sec:Tdual}

So far, we have presented three T-dual backgrounds. The corresponding string coordinates can be related thanks to the T-duality relations derived in \eqref{Tdcoordrel}, so let us first give the latter explicitly. We use the rescaled quantities in these relations (see the discussion around table \ref{tab:resc} in appendix \ref{ap:not}), in particular the target space fields of table \ref{tab:rescandH}, and obtain
\bea
& \mbox{T-d. along } \iota=1: \partial_{\tau} X^{2,3}=\partial_{\tau}  Y^{2,3},\  \partial_{\sigma} X^{2,3}=\partial_{\sigma}Y^{2,3},\ \mbox{and}\nn\\
& \qquad \qquad \begin{array}{r|c|l}
 \partial_{\tau} X^1=\partial_{\sigma} Y^1-HY^3\partial_{\sigma} Y^2 & \Longleftrightarrow & \partial_{\tau} Y^1=\partial_{\sigma} X^1+HX^3\partial_{\tau} X^2  \\
 \partial_{\sigma} X^1=\partial_{\tau} Y^1-HY^3 \partial_{\tau} Y^2 & \mbox{(all order in $H$)} & \partial_{\sigma} Y^1=\partial_{\tau} X^1+HX^3 \partial_{\sigma} X^2
\end{array} \label{eq:TdcoordXY}
\eea
\bea
& \mbox{T-d. along } \iota=2: \partial_{\tau} Y^{1,3}=\partial_{\tau}  Z^{1,3},\ \partial_{\sigma} Y^{1,3}=\partial_{\sigma}Z^{1,3},\ \mbox{and}\nn\\
& \qquad \qquad \begin{array}{r|c|l}
 \partial_{\tau} Y^2=\partial_{\sigma} Z^2+HZ^3\partial_{\tau} Z^1 & \Longleftrightarrow & \partial_{\tau} Z^2=\partial_{\sigma} Y^2-HY^3\partial_{\sigma} Y^1  \\
 \partial_{\sigma} Y^2=\partial_{\tau} Z^2+HZ^3 \partial_{\sigma} Z^1 & \mbox{(up to ${\cal O}(H^2)$)} & \partial_{\sigma} Z^2=\partial_{\tau} Y^2-HY^3 \partial_{\tau} Y^1
\end{array} \label{eq:TdcoordYZ}
\eea
and by composition
\bea
& \mbox{T-d. along $1$ and $2$}: \partial_{\tau} X^{3}=\partial_{\tau}  Z^{3},\ \partial_{\sigma} X^{3}=\partial_{\sigma}Z^{3},\ \mbox{and}\nn\\
& \qquad \qquad \begin{array}{r|c|l}
 \partial_{\tau} X^1=\partial_{\sigma} Z^1-HZ^3\partial_{\tau} Z^2 &  & \partial_{\tau} Z^1=\partial_{\sigma} X^1+HX^3\partial_{\tau} X^2  \\
 \partial_{\sigma} X^1=\partial_{\tau} Z^1-HZ^3 \partial_{\sigma} Z^2 & \Longleftrightarrow & \partial_{\sigma} Z^1=\partial_{\tau} X^1+HX^3 \partial_{\sigma} X^2 \\
 \partial_{\tau} X^2=\partial_{\sigma} Z^2+HZ^3\partial_{\tau} Z^1 & \mbox{(up to ${\cal O}(H^2)$)} & \partial_{\tau} Z^2=\partial_{\sigma} X^2-HX^3\partial_{\tau} X^1  \\
 \partial_{\sigma} X^2=\partial_{\tau} Z^2+HZ^3 \partial_{\sigma} Z^1 &  & \partial_{\sigma} Z^2=\partial_{\tau} X^2-HX^3 \partial_{\sigma} X^1
\end{array} \label{eq:TdcoordXZ}
\eea

We find that the derivatives of the third coordinate always match, which is consistent with this coordinate being unaffected by T-duality along ${\cal X}^1$ and ${\cal X}^2$. It means that we have a base circle in the geometry which remains invariant under T-dualities of a two-torus that is fibered over it. This equality of the derivatives of the third coordinate indicates that $X^3, Y^3, Z^3$ could be identified, up to a possible difference in the center of mass position constant. However, we mentioned previously the need for the exact identification, so that the target space fields are T-dual. We refine this requirement in what follows, and will only identify the zeroth order in $H$, i.e. $X_0^3=Y_0^3=Z_0^3$. Doing so we ignore the freedom in the zeroth order center of mass position.\\

Let us now make use of the relations \eqref{eq:TdcoordXY} to investigate the equations of motion for the two geometric T-dual backgrounds. Under T-duality along ${\cal X}^1$, the equations of motion (e.o.m.) of the two coordinates that are not dualised match:
\bea
X^2 \ {\rm e.o.m.}\ \eqref{Aeom} & \Leftrightarrow Y^2 \ {\rm e.o.m.}\ \eqref{eqY2} \label{eq:eomTd1}\\
X^3 \ {\rm e.o.m.}\ \eqref{Aeom} & \Leftrightarrow Y^3 \ {\rm e.o.m.}\ \eqref{eqY3} \ , \label{eq:eomTd2}
\eea
as can be seen by simply using \eqref{eq:TdcoordXY}. Here, we consider the equations of motion valid to all order in $H$, so those relations are always valid. Again, this is consistent with T-duality only affecting the coordinate that is dualised. Interestingly, the $X^1$ e.o.m. \eqref{Aeom} and $Y^1$ e.o.m. \eqref{eqY1} do not lead to one another when using the T-duality relations \eqref{eq:TdcoordXY}; doing so rather makes them automatically satisfied to all orders (one obtains trivial identities) \cite{Duff:1989tf}. To get these e.o.m., one only needs the T-duality relations \eqref{eq:TdcoordXY} and then considers the (trivial) equalities
\bea
& 0=\del_{\sigma}(\del_{\tau} X^1) - \del_{\tau}(\del_{\sigma} X^1) \Rightarrow Y^1 \ {\rm e.o.m.}\ \eqref{eqY1} \ , \label{eq:trivialeomXY}\\
& 0=\del_{\sigma}(\del_{\tau} Y^1) - \del_{\tau}(\del_{\sigma} Y^1) \Rightarrow X^1 \ {\rm e.o.m.}\ \eqref{Aeom} \ . \label{eq:trivialeom}
\eea
To conclude, given the T-duality relations, the equations of motion in one background can be obtained from those of the T-dual background, together with trivial constraints.

Likewise, one can easily show that the closed string boundary conditions for $X$, given in \eqref{eq:bdyX}, and $Y$, given in \eqref{bdy1}, \eqref{bdy2}, and \eqref{bdy3}, are mapped onto each other by the T-duality transformation \eqref{eq:TdcoordXY}, up to the precise value of the winding constants. In particular, one can notice that the non-trivial Maurer-Cartan one-form of the twisted torus is precisely the quantity entering in these T-duality relations; its global-definedness is then mapped.\\

These properties will play an important role in what follows. The map of equations of motion and boundary conditions in not dualised directions implies that the classical string solutions in these directions can as well be mapped. This requires the T-duality relations to hold, and those are also important for the map of the dualised direction. Verifying explicitly these T-duality relations may bring new constraints, but will then guarantee that starting from a string solution, one gets a solution after T-duality. We will investigate these important features in more details for the geometric T-dual backgrounds, and use this same idea as a starting point in our analysis of the non-geometric background in section \ref{sec:nongeo}.

At linear order in $H$, the world-sheet equations of motion for the non-geometric background should locally be the same as those of the torus with $H$-flux, up to a sign on $H$, as can be seen on the target space fields in table \ref{tab:rescandH}. We however expect the classical solution to differ, and this can be understood as a consequence of the different boundary conditions. The latter encode the global aspects, which are known to differ between geometric and non-geometric backgrounds. The relation between the global aspects and the boundary conditions is also discussed in appendix \ref{sec:monod} when studying monodromies. Here, one can verify explicitly, by looking at the T-duality relations \eqref{eq:TdcoordXZ}, that the boundary conditions cannot be the same between these two backgrounds.

%%%%%%%%%%%%%%%%%%%%%%%%%%%%
\subsection{Canonical commutation relations and T-duality}\label{sec:quantgen}

In this section, we turn to quantum aspects. We initiate the canonical quantization by presenting the general canonical commutation relations for the string, and their explicit $H$ expansions, for the two geometric backgrounds at hand. We then study how T-duality relates the canonical commutators of the two backgrounds, and find that the information of these commutators in one background can be obtained by studying the commutators in a T-dual one. This will be useful in the study of the non-geometric situation in section \ref{sec:nongeo}.\\

In any geometric background, the canonical commutation relations are\footnote{We use right away the rescaled quantities, see the discussion around table \ref{tab:resc} in appendix \ref{ap:not}.}
\bea
[{\cal X}^{\mu}(\tau,\sigma),{\cal X}^{\nu}(\tau,\sigma')] & = 0 \label{XX} \\
[{\cal P}_{\mu}(\tau,\sigma),{\cal P}_{\nu}(\tau,\sigma')] & = 0 \label{PP} \\
[{\cal X}^{\mu}(\tau,\sigma),{\cal P}_{\nu}(\tau,\sigma')] & = \i \ \delta^{\mu}_{\nu}\ \delta(\sigma-\sigma') \ ,\label{XP}
\eea
where
\beq
{\cal P}_{\mu}
\equiv \frac{\delta \L}{\delta \del_{\tau} {\cal X}^{\mu}}
=\frac{1}{\pi} \left(G_{\mu\nu}({\cal X}) \del_{\tau} {\cal X}^{\nu} + B_{\mu\nu}({\cal X}) \del_{\sigma} {\cal X}^{\nu} \right) \ ,\label{canmom}
\eeq
is the canonical momentum. Here the Lagrangian $\L$ is read from the action $S=\int_\Sigma d^2\sigma \ \L$ in \eqref{action}, and we follow the conventions given there. Importantly, ${\cal P}_{\mu}$ depends on the background fields and thus differ between the torus with $H$-flux and the twisted torus backgrounds. To compute ${\cal P}_{\mu}$ and the canonical commutators, we use the target space fields in table \ref{tab:rescandH}, and expand the coordinates fields to first order in $H$ as
\beq
{\cal X}^{\mu}(\tau,\sigma) = {\cal X}_0^{\mu}(\tau,\sigma) + H {\cal X}_H^{\mu}(\tau,\sigma) \ ,
\eeq
where ${\cal X}_0$ (${\cal X}_H$) solves the equation of motion and boundary conditions to zeroth (first) order in $H$. The full solutions will be presented in section \ref{sec:solutions}, and we will use this to start an explicit canonical quantization of the string on the twisted torus in section \ref{sec:quanttor}.

\paragraph{Zeroth order}

At zeroth order in $H$, the torus with $H$-flux and the twisted torus reduce to a fluxless torus background. For both of them, the rescaled background metric then boils down to $\eta_{\mu \nu}$, $B$ vanishes and the canonical commutation relations become simply
\bea
[{\cal X}_0^{\mu}(\tau,\sigma),{\cal X}_0^{\nu}(\tau,\sigma')] & = 0 \label{X0X0} \\
[\del_{\tau} {\cal X}_0^{\mu}(\tau,\sigma),\del_{\tau} {\cal X}_0^{\nu}(\tau,\sigma')] & = 0 \label{dtX0dtX0} \\
[{\cal X}_0^{\mu}(\tau,\sigma),\del_{\tau} {\cal X}_0^{\nu}(\tau,\sigma')] & = \i\pi \ \eta^{\mu\nu}\ \delta(\sigma-\sigma')\ . \label{YdtY0}
\eea
These are the standard relations for the free string, as expected.

\paragraph{First order}

To start with, the commutator of two coordinates still vanishes as given by \eqref{XX}, but this translates into
\beq
0=[{\cal X}^{\mu}(\tau,\sigma),{\cal X}^{\nu}(\tau,\sigma')]|_H = H\ [{\cal X}_0^{\mu}(\tau,\sigma),{\cal X}_H^{\nu}(\tau,\sigma')] + H\ [{\cal X}_H^{\mu}(\tau,\sigma),{\cal X}_0^{\nu}(\tau,\sigma')] \ ,\label{XXH}
\eeq
where by $|_H$ we mean exactly the $H$-order term (it does not contain the zeroth order term). This commutator relation holds for the two geometric backgrounds.

We now turn to the commutators that involve the canonical momentum, and start with the torus with $H$-flux. From \eqref{XP}, using \eqref{canmom}, the zeroth order commutators, and the target space fields, it is easy to deduce
\beq
[X^{\mu}(\tau,\sigma),\del_{\tau} X^{\nu}(\tau,\sigma')]|_H = 0 \ .\label{XdtXH}
\eeq
It is less straightforward to deduce the commutator of two $\del_{\tau}X^{\mu}$ from the one involving two canonical momenta \eqref{PP}. Using again \eqref{canmom}, the zeroth order commutators, and the target space fields, one finds that the non-trivial commutators are
\bea
& [\del_{\tau} X^1(\tau,\sigma),\del_{\tau} X^2(\tau,\sigma')]|_H = \i \pi H\ \left(X^3_0(\tau,\sigma')- X^3_0(\tau,\sigma) \right)\ \del_{\sigma} \delta(\sigma-\sigma') \label{dtX1dtX2H}\\
& [\del_{\tau} X^3(\tau,\sigma),\del_{\tau} X^1(\tau,\sigma')]|_H = \i \pi H\ \delta(\sigma-\sigma')\ \del_{\sigma'}X^2_0(\tau,\sigma') \label{dtX3dtX1H}\\
& [\del_{\tau} X^3(\tau,\sigma),\del_{\tau} X^2(\tau,\sigma')]|_H = -\i \pi H\ \delta(\sigma-\sigma')\ \del_{\sigma'}X^1_0(\tau,\sigma') \ .\label{dtX3dtX2H}
\eea

For the twisted torus we have ${\cal P}_{\mu} = \frac{1}{\pi} G_{\mu\nu}(Y) \del_{\tau} Y^{\nu}$, so by multiplying \eqref{XP} on the left by $G^{\rho\nu}(Y)(\tau,\sigma')$ and using \eqref{XX}, we get
\bea
[Y^{\mu}(\tau,\sigma),\del_{\tau} Y^{\nu}(\tau,\sigma')] & = \i\pi \ \delta(\sigma-\sigma')\ G^{\mu\nu}(Y) (\tau,\sigma') \ .\label{XP2}
\eea
The above inverse metric, at first order in $H$, can be obtained from the one in table \ref{tab:rescandH} by only changing the sign of the off-diagonal components. From this we obtain one non-zero commutator (for any other combination of coordinate indices, \eqref{XP2} at first order is zero)
\bea
& [Y^{1}(\tau,\sigma),\del_{\tau} Y^{2}(\tau,\sigma')]|_H=[Y^2(\tau,\sigma),\del_{\tau} Y^1(\tau,\sigma')]|_H = \i \pi H \ \delta(\sigma - \sigma') \ Y^3_0 (\tau,\sigma') \ .\label{Y1dY2H}
\eea
Finally, to relate \eqref{PP} to the commutator of two $\del_{\tau} Y^{\mu}$, we first consider the commutator of two $G^{\mu\rho}\ {\cal P}_{\rho}$. It is more involved than \eqref{PP}, because of the dependence of the metric on $Y$. Using the identity $[AB,C]=A[B,C]+[A,C]B$, together with \eqref{XP} and the explicit form of the inverse metric, we deduce
\bea
& [\del_{\tau} Y^3(\tau,\sigma),\del_{\tau} Y^1(\tau,\sigma')]|_H = -\i \pi H\ \delta(\sigma - \sigma') \ \del_{\tau} Y_0^2 (\tau,\sigma') \ ,\label{dtYdtYH31}\\
& [\del_{\tau} Y^3(\tau,\sigma),\del_{\tau} Y^2(\tau,\sigma')]|_H = -\i \pi H\ \delta(\sigma - \sigma') \ \del_{\tau} Y_0^1 (\tau,\sigma') \ ,\label{dtYdtYH32}\\
\textrm{all}\ \textrm{other}\ \ & [\del_{\tau} Y^{\mu}(\tau,\sigma),\del_{\tau} Y^{\nu}(\tau,\sigma')]|_H = 0 \ .\label{dtYdtYH}
\eea
In addition, we get using \eqref{Y1dY2H}, or $\del_{\sigma} \eqref{XP}$,
\beq
[\del_{\sigma} Y^{1}(\tau,\sigma),\del_{\tau} Y^{2}(\tau,\sigma')]|_H=[\del_{\sigma} Y^2(\tau,\sigma),\del_{\tau} Y^1(\tau,\sigma')]|_H = \i \pi H \ Y^3_0 (\tau,\sigma') \ \del_{\sigma}\delta(\sigma - \sigma') \ ,\label{dsY1dtY2H}
\eeq
that we combine with \eqref{XXH} and \eqref{dtYdtYH}, for future convenience, to
\bea
[\del_{\sigma_{\epsilon_1}}Y^1(\tau,\sigma),\del_{\sigma_{\epsilon_2}'} Y^2(\tau,\sigma')]|_H & = [\del_{\sigma_{\epsilon_1}}Y^2(\tau,\sigma),\del_{\sigma_{\epsilon_2}'} Y^1(\tau,\sigma')]|_H \label{dsY1dsY2H}\\
& = \frac{\i \pi}{4} H \ \Big(\epsilon_1\ Y^3_0 (\tau,\sigma') +\epsilon_2\ Y^3_0 (\tau,\sigma) \Big) \ \del_{\sigma} \delta(\sigma - \sigma') \ ,\nn
\eea
introducing the notation ${\epsilon}=\pm1$ so that $\sigma_{\epsilon} = \sigma_{\pm} $.\\

With these canonical commutators at order $H$ at hand, we can investigate how they are related under T-duality along ${\cal X}^1$. Since the T-duality only relates the derivatives of the coordinates, it is primarily the $\sigma$ or $\sigma'$ derivatives of these commutators that are mapped. More precisely, provided the zeroth order commutation relations \eqref{X0X0} - \eqref{YdtY0} and the T-duality relations \eqref{eq:TdcoordXY}, one shows that one set of the following commutators gives the other one
\begin{center}
\begin{tabular}{l|c|l}
 $[\del_{\sigma} X^{\mu}(\tau,\sigma),\del_{\sigma'} X^{\nu}(\tau,\sigma')]|_H $ &  & $[\del_{\sigma} Y^{\mu}(\tau,\sigma),\del_{\sigma'} Y^{\nu}(\tau,\sigma')]|_H $\\
 $[\del_{\sigma} X^{\mu}(\tau,\sigma),\del_{\tau} X^{\nu}(\tau,\sigma')]|_H $  & $\Longleftrightarrow$ & $[\del_{\sigma} Y^{\mu}(\tau,\sigma),\del_{\tau} Y^{\nu}(\tau,\sigma')]|_H $\\
 $[\del_{\tau} X^{\mu}(\tau,\sigma),\del_{\tau} X^{\nu}(\tau,\sigma')]|_H $ & (at order $H$) & $[\del_{\tau} Y^{\mu}(\tau,\sigma),\del_{\tau} Y^{\nu}(\tau,\sigma')]|_H $
\end{tabular}
\end{center}
or in other words
\begin{center}
\begin{tabular}{r|c|l}
 $\del_{\sigma} \del_{\sigma'} \eqref{XXH}$ &  & $\del_{\sigma} \del_{\sigma'} \eqref{XXH}$ \\
 $\del_{\sigma} \eqref{XdtXH}$ & $\Longleftrightarrow$ & $\del_{\sigma} \eqref{XP2}$ \\
 $\eqref{dtX1dtX2H}, \ \eqref{dtX3dtX1H}, \ \eqref{dtX3dtX2H}$ & (at order $H$) & $\eqref{dtYdtYH31}, \ \eqref{dtYdtYH32}, \ \eqref{dtYdtYH}$
\end{tabular}
\end{center}
For instance, one reproduces \eqref{dtX1dtX2H} as follows
\bea
[\del_{\tau} X^1(\tau,\sigma),\del_{\tau} X^2(\tau,\sigma')]|_H & = [\del_{\sigma} Y^1(\tau,\sigma),\del_{\tau} Y^2(\tau,\sigma')]|_H - H\ [Y_0^3 \del_{\sigma} Y_0^2 (\tau,\sigma), \del_{\tau} Y_0^2(\tau,\sigma') ] \nn\\
& = \i \pi H\ \left(Y^3_0(\tau,\sigma')- Y^3_0(\tau,\sigma) \right)\ \del_{\sigma} \delta(\sigma-\sigma') \  {\rm using}\ \del_{\sigma} \eqref{XP2}, \ \eqref{YdtY0}\nn\\
& = \i \pi H\ \left(X^3_0(\tau,\sigma')- X^3_0(\tau,\sigma) \right)\ \del_{\sigma} \delta(\sigma-\sigma') \ .\nn
\eea

To conclude, given the relations between coordinates in T-dual backgrounds, we can use the information in one background to compute commutators in another. This will be useful when we analyse the non-geometric situation in section \ref{sec:nongeo}.

%%%%%%%%%%%%%%%%%%%%%%%%%%%%
\section{Analysis of the geometric backgrounds}\label{sec:geo}

In this section, we analyse the torus with $H$-flux and the twisted torus in more detail. For these backgrounds, we find the solutions to the equations of motion and boundary conditions to linear order in $H$, and discuss how they are related by T-duality. Subsequently, we turn to the canonical quantization of the twisted torus. The classical modes of its solution are promoted to operators, and we derive those commutation relations that are needed for the analysis of the non-geometric set-up in section \ref{sec:nongeo}.

\subsection{Solutions for the coordinate fields}\label{sec:solutions}

The equations of motion and boundary conditions for the coordinate fields were given in sections \ref{subsubsec:Htorus} and \ref{subsubsec:Ttorus}. Using the dilute flux approximation, we will now find solutions to these equations up to linear order in $H$. Our presentation will be fairly brief, as the (lengthy) computations are straightforward and the analyses of the two T-dual situations are very similar (although the respective solutions differ).

We start our analysis by examining the boundary conditions, i.e. \eqref{eq:bdyX} for the torus with $H$-flux, and \eqref{bdy1}-\eqref{bdy3} for the twisted torus. These boundary conditions show that while the coordinates are not periodic in $\sigma$, the following functions are
\beq \nn
X^{\mu} - N_X^{\mu} \sigma
\quad , \quad
Y^{2,3} - N^{2,3} \sigma
\quad , \quad
Y^1 -N^1 \sigma -H\left( N^3  \sigma (Y^2 - N^2 \sigma) - \frac{1}{2} N^3  N^2 \sigma (2\pi - \sigma) \right)\ .
\eeq
Therefore, all these functions can be expanded as Fourier series in $\sigma$, with $\tau$-dependent expansion coefficients. We can develop these expansion coefficients order by order in $H$, so that at linear order, we get for the torus with $H$-flux
\beq
X^{\mu}(\tau,\sigma) = N_X^{\mu} \sigma + \sum_{n \in \Z} b^{\mu}_{Xn} (\tau) e^{-\i n \sigma} + H \Big( \sum_{n \in \Z} c^{\mu}_{Xn} (\tau) e^{-\i n \sigma} \Big) \ ,\label{eq:formX}
\eeq
and, for the twisted torus,
\bea
Y^1(\tau,\sigma) =&~N^1 \sigma + \sum_{n \in \Z} b^1_{n} (\tau) e^{-\i n \sigma} \nn\\
& + H \Big( N^3 \sigma (Y^2 - N^2 \sigma) - \frac{1}{2} N^3 N^2 \sigma (2\pi - \sigma) + \sum_{n \in \Z} c^1_{n} (\tau) e^{-\i n \sigma} \Big) \label{form1}\\
Y^{2,3}(\tau,\sigma) =&~N^{2,3} \sigma + \sum_{n \in \Z} b^{2,3}_{n} (\tau) e^{-\i n \sigma} + H \Big( \sum_{n \in \Z} c^{2,3}_{n} (\tau) e^{-\i n \sigma} \Big) \ ,\label{form23}
\eea
where the $b_X, b$ and $c_X, c$ are $H$-independent functions of $\tau$. We will determine the coefficients of these series by inserting \eqref{eq:formX}, \eqref{form1} and \eqref{form23} into the equations of motion \eqref{Aeom} and \eqref{eq:eomY} respectively. This will lead to solutions of the form
\beq
{\cal X}^{\mu}(\tau,\sigma) = {\cal X}_0^{\mu}(\tau,\sigma) + H {\cal X}_H^{\mu}(\tau,\sigma)  + {\cal O}(H^2)\; ,
\eeq
where ${\cal X}_0$ is the solution to the equation of motion at ${\cal O}(H^0)$, and ${\cal X}_H$ is the solution at ${\cal O}(H)$.

\paragraph{Twisted torus} Let us now focus on the string in the twisted torus background. The analysis for the torus with $H$-flux proceeds in the same way, and we will only present the result below.  At ${\cal O}(H^0)$ the equation of motion and boundary conditions are just those for the free string, and the solution is given by (with $\sigma_{\pm}=\tau \pm \sigma$)
\beq
Y^{\mu}_0=y^{\mu}+p^{\mu} \tau + N^{\mu} \sigma +\frac{\i}{2}\sum_{n\neq0}\frac{1}{n} \left( \widetilde{\alpha}_n^{\mu} e^{-\i n\sigma_+} + \alpha_n^{\mu} e^{-\i n\sigma_-} \right) \ . \label{eq:sol0}
\eeq
For future convenience, let us decompose these into left- and right-moving parts
\bea \label{eq:sol0LR}
Y_{0L}^{\mu}&=y_L^{\mu}+p_L^{\mu} \sigma_+ +\frac{\i}{2}\sum_{n\neq0}\frac{1}{n}\widetilde{\alpha}_n^{\mu} e^{-\i n \sigma_+} \quad \ , \quad
Y_{0R}^{\mu} &= y_R^{\mu}+p_R^{\mu} \sigma_- +\frac{\i}{2}\sum_{n\neq0}\frac{1}{n}\alpha_n^{\mu} e^{-\i n \sigma_-} \ ,\\
 Y^{\mu}_0 &= Y_{0L}^{\mu}+Y_{0R}^{\mu}\ , \ \tilde{Y}^{\mu}_0=Y_{0L}^{\mu}-Y_{0R}^{\mu} \  \quad \ , \quad
{\rm and} & y^{\mu} = y^{\mu}_L+y^{\mu}_R \ , \ \tilde{y}^{\mu}=y^{\mu}_L-y^{\mu}_R \; \;  \;   ,\nn\\
& &  p^{\mu}=p^{\mu}_L+p^{\mu}_R \ , \ N^{\mu}=p^{\mu}_L-p^{\mu}_R \ .\nn
\eea
The division of $y^{\mu}$ is conventional, and ensures that $y^{\mu}, \tilde{y}^{\mu}$ are dual to $p^{\mu}, N^{\mu}$ respectively.

Proceeding to linear order in $H$, \eqref{eq:eomY} gives\footnote{This order by order method decomposes the non-linear equations of motion for the twisted torus as two linear differential equations: the wave equation and \eqref{eqH1}, where the latter has a non-zero right-hand side. This is a great simplification, since generic properties of such equations, as the decomposition into homogeneous and particular solutions, can thus be used.}
\beq
\partial_\alpha\partial^\alpha Y_H^{\mu} = {\theta^{\mu}}_{\nu\rho}\ \partial_\alpha Y_0^{\nu} \partial^\alpha Y_0^{\rho} \ , \label{eqH1}
\eeq
where we recall that ${\theta^1}_{23}={\theta^2}_{13}=-{\theta^3}_{12}=1$, and all other components are zero. We define similarly  ${\lambda^{\mu}}_{\nu\rho}$ as ${\lambda^1}_{23}=1$, and all other components are zero. With these definitions the equations at order $H$ given by \eqref{eqH1} become, using \eqref{form1} and \eqref{form23},
\bea
\partial_\alpha\partial^\alpha \left(\sum_{n \in \Z} c^{\mu}_{n} (\tau) e^{-\i n \sigma} \right) &= {\theta^{\mu}}_{\nu\rho}\ \partial_\alpha Y_0^{\nu} \partial^\alpha Y_0^{\rho} \label{eqH2} \\
& - {\lambda^{\mu}}_{23}\  \partial_\alpha\partial^\alpha \left(N^3 \sigma (Y_0^2 - N^2 \sigma) - \frac{1}{2} N^3 N^2 \sigma (2\pi - \sigma) \right) \ , \nn
\eea
Solving this equation for the Fourier coefficients $c^{\mu}_{n}$, one finds the most general solutions to \eqref{eqH1} and the boundary conditions \eqref{bdy1}-\eqref{bdy3}. After some non-trivial rearranging, the solutions take the form
\bea
Y^{\mu}_H (\tau, \sigma) =&~y_H^{\mu} + p_H^{\mu} \ \tau +\frac{\i}{2}\sum_{n\neq0}\frac{1}{n} \left( \tg_n^{\mu} e^{-\i n\sigma_+} + \g_n^{\mu} e^{-\i n\sigma_-} \right) \label{finalsol}\\
& + {\theta^{\mu}}_{\nu\rho}\ (p^{\rho} p^{\nu} -N^{\rho} N^{\nu}) \ \frac{\tau^2}{2} \nn\\
& +{\theta^{\mu}}_{\nu\rho}\ \frac{1}{2}\ \tau\ \left(p^{\rho} Y_0^{\nu}|_{\Sigma} - N^{\rho} \tilde{Y}_0^{\nu}|_{\Sigma} + p^{\nu} Y_0^{\rho}|_{\Sigma} - N^{\nu} \tilde{Y}_0^{\rho}|_{\Sigma} \right) \nn\\
& -{\theta^{\mu}}_{\nu\rho}\ \frac{1}{4} \left(\tilde{Y}_0^{\nu}|_{\Sigma} \tilde{Y}_0^{\rho}|_{\Sigma} - Y_0^{\nu}|_{\Sigma} Y_0^{\rho}|_{\Sigma} \right) \nn\\
& + {\lambda^{\mu}}_{23}\ N^3 \left( N^2 \ \frac{\tau^2}{2} + \tau\ \tilde{Y}_0^{2}|_{\Sigma} + \sigma (Y_0^2 - N^2 \sigma) - \frac{1}{2} N^2 \sigma (2\pi - \sigma) \right) \nn\ ,
\eea
where $y_H^{\mu}, p_H^{\mu}, \tg_n^{\mu}$ and $\g_n^{\mu}$ are arbitrary constants and we denote
\beq
Y_0^{\mu}|_{\Sigma}=\frac{\i}{2}\sum_{n\neq0}\frac{1}{n} \left( \widetilde{\alpha}_n^{\mu} e^{-\i n\sigma_+} + \alpha_n^{\mu} e^{-\i n\sigma_-} \right) \ . \label{XSigma}
\eeq

The expression \eqref{finalsol} makes it clear that the boundary conditions are satisfied. Indeed, all terms but the ${\lambda^{\mu}}_{23}$ ones are periodic in $\sigma$, and the (last two) ${\lambda^{\mu}}_{23}$ terms give precisely the boundary conditions \eqref{bdy1}-\eqref{bdy3}. It is less obvious that this expression solves the equation of motion \eqref{eq:eomY}. However, one can actually rewrite the solution as
\beq
Y^{\mu}_H (\tau, \sigma) = -\frac{{\theta^{\mu}}_{\nu\rho}}{4} \left(\tilde{Y}_0^{\nu} \tilde{Y}_0^{\rho} - Y_0^{\nu} Y_0^{\rho} \right) + f^{\mu}_L (\sigma_+) + f^{\mu}_R (\sigma_-) \ ,\label{solY0Y0}
\eeq
where the left- and right-moving functions  $f_{L/R}$ can be computed straightforwardly. In addition, by decomposing on left and right movers and using $2\partial_{\sigma_{\pm}} = \partial_{\tau} \pm \partial_{\sigma}$, one finds
\beq
-\frac{1}{4} \del_{\alpha} \del^{\alpha} (\tilde{Y}_0^{\nu} \tilde{Y}_0^{\rho} - Y_0^{\nu} Y_0^{\rho}) = \del_{\alpha} Y_0^{\nu} \del^{\alpha} Y_0^{\rho} \ .
\eeq
The expression \eqref{solY0Y0} then makes it clear that $Y^{\mu}_H$ solves the equation of motion at order $H$. The price to pay  for this reformulation is that the rewritten solution no longer manifestly satisfies the boundary conditions. Note that the reformulation \eqref{solY0Y0} requires a split of $y^{\mu}_H$ into $y^{\mu}_{H\,L,R}$, just as for the free string. We will use this decomposition in appendix \ref{sec:fixcom}.

\paragraph{Torus with $H$-flux} To complete this section, let us record the solution at zeroth and linear order in $H$ for a string propagating in a torus with $H$-flux. Using the same methods as above, we find
\bea
X^{\mu}_0=&~x^{\mu}+p_X^{\mu} \tau + N_X^{\mu} \sigma +\frac{\i}{2}\sum_{n\neq0}\frac{1}{n} \left( \widetilde{\alpha}_{Xn}^{\mu} e^{-\i n\sigma_+} + \alpha_{Xn}^{\mu} e^{-\i n\sigma_-} \right) \ ,\label{eq:sol0X} \\
X^{\mu}_H (\tau, \sigma) =&~x_H^{\mu} + p_{HX}^{\mu} \ \tau +\frac{\i}{2}\sum_{n\neq0}\frac{1}{n} \left( \tg_{Xn}^{\mu} e^{-\i n\sigma_+} + \g_{Xn}^{\mu} e^{-\i n\sigma_-} \right) \label{Xfinalsol}\\
& - {\epsilon^{\mu}}_{\nu\rho}\left(  p_X^{\rho} N_X^{\nu} \ \frac{\tau^2}{2} +
 \frac{1}{2}\ \tau\ \left(N_X^{\nu} X_0^{\rho}|_{\Sigma} -  p_X^{\nu} \tilde{X}_0^{\rho}|_{\Sigma} \right) + \frac{1}{4}\tilde{X}_0^{\nu}|_{\Sigma} X_0^{\rho}|_{\Sigma}  \right) \nn\ ,
\eea
where $x^{\mu}, p_{X}^{\mu}, N_X^{\mu}, \widetilde{\alpha}_{Xn}^{\mu}, \alpha_{Xn}^{\mu}, x_H^{\mu}, p_{HX}^{\mu}, \tg_{Xn}^{\mu}$ and $\g_{Xn}^{\mu}$ are arbitrary constants.\footnote{A solution to the equation of motion \eqref{Aeom} has previously been presented in \cite{Blumenhagen:2011ph}. This solution differs from the one presented here, in that it fails to satisfy the boundary condition \eqref{eq:bdyX}. For the sake of clarity, let us also mention that since we are considering an ${\cal O}(H)$ correction to the coordinate solutions, we avoid a restriction on the string zero modes that is present in \cite{Blumenhagen:2011ph}.}

\paragraph{Constraints from T-duality}
Given the solutions for $X^{\mu}$ and $Y^{\mu}$, we can now use the T-duality relations in section \ref{sec:Tdual} to relate their expansion coefficients. At zeroth order, the T-duality rules \eqref{eq:TdcoordXY} are obeyed if we match
\bea
p^1_X &= N^1 \; , \;
N^1_X = p^1 \; , \;
p^{2,3}_X = p^{2,3} \; , \;
N^{2,3}_X = N^{2,3} \; , \\
\alpha^1_{Xn} &= -\alpha^1_n \; , \;
\widetilde{\alpha}^1_{Xn} = \widetilde{\alpha}^1_n \; , \;
\alpha^{2,3}_{Xn} = \alpha^{2,3}_n \; , \;
\widetilde{\alpha}^{2,3}_{Xn} = \widetilde{\alpha}^{2,3}_n \; . \;
\eea
This simple match is expected from the T-duality relations \eqref{eq:TdcoordXY}. Since $x^{\mu}$ and $y^{\mu}$ drop out when taking derivatives, we seem to be free to choose them independently. However, using the decomposition into left and right movers \eqref{eq:sol0LR} in conjunction with the fact that the canonical commutators \eqref{X0X0}-\eqref{YdtY0} hold in both frames, we have at the level of commutators
\beq
x^1 = \tilde{y}^1  \; , \; x^{2,3} = y^{2,3} \; . \label{eq:Magda}
\eeq
In other words, we must identify the commutation relations of these operators. This reproduces the standard T-dual solutions for the free string, as expected.

We proceed to first order in $H$, and directions 2 and 3 (i.e. $X^{2,3}_H$ and $Y^{2,3}_H$). Just as at zeroth order, the equations of motion and the boundary conditions for these coordinate fields are mapped to each other by T-duality, see \eqref{eq:eomTd1}-\eqref{eq:eomTd2}. We thus expect the solutions to match by a simple identification. This is indeed the case, and the identification is
\beq
p_{HX}^{2,3} = p_H^{2,3}  \; , \;
\g^{2,3}_{Xn} = \g^{2,3}_n  \; , \;
\tg^{2,3}_{Xn} = \tg^{2,3}_n  \; . \;
\eeq
Again, $x_H^{\mu}$ and $y_H^{\mu}$ drop out when taking derivatives. Inspired by the zeroth order relations, $x^{2,3}_H = y^{2,3}_H$ is a natural choice.

For  $X^{1}_H$ and $Y^{1}_H$, the situation is more involved, since the equations of motion do not map to each other, but rather to trivial conditions, 
see \eqref{eq:trivialeomXY}-\eqref{eq:trivialeom}. Nevertheless, by imposing the T-duality relations \eqref{eq:TdcoordXY}, we can solve for one set of expansion coefficients in terms of the other. We get the following conditions
\bea
\tg_{Xn}^1 &= \;\;\; \left( \tg^1_n + \frac{\i}{n} (p^3_L \widetilde{\alpha}^2_n -p^2_L \widetilde{\alpha}^3_n) -\frac{1}{2} y_3 \widetilde{\alpha}^2_n - \frac{\i}{2} \sum_{m\neq0,n} \frac{1}{m} \widetilde{\alpha}^3_m \widetilde{\alpha}^2_{n-m} \right) \label{eq:fixg1Xm+}\\
\g_{Xn}^1 &= -\left(\g^1_n + \frac{\i}{n} (p^3_R \alpha^2_n -p^2_R {\alpha}^3_n) - \frac{1}{2} y_3 {\alpha}^2_n - \frac{\i}{2} \sum_{m\neq0,n} \frac{1}{m} \alpha^3_m {\alpha}^2_{n-m}\right) \; \label{eq:fixg1Xm-} \\
p_{HX}^1 &=  \left( N^3 y^2 - N^2 y^3 - \pi N^2 N^3 \right) - \frac{\i}{4} \sum_{n\neq0} \frac{1}{n} \left( \widetilde{\alpha}^3_n \widetilde{\alpha}^2_{-n} -{\alpha}^3_n {\alpha}^2_{-n}\right) \label{eq:fixpHX1} \\
p_H^1 &=  y^3 p^2 + \frac{\i}{4} \sum_{n\neq0} \frac{1}{n} \left( \widetilde{\alpha}^3_n \widetilde{\alpha}^2_{-n} +{\alpha}^3_n {\alpha}^2_{-n}\right) \; . \label{eq:fixpH1}
\eea
As expected, the first three of these equations relate modes in the $H$-flux background to modes for the twisted torus solutions. On the contrary, the last condition is a constraint that acts exclusively on the zero modes in the twisted torus set-up. The reason for this is that we have imposed, in both situations, that there is no $\mathcal{O}(H)$ correction to the winding numbers (see the boundary conditions). Since momentum and winding modes are mapped by T-duality, it is clear that imposing a constraint on the winding mode in one frame, will restrict the T-dual momentum mode, and thus the last constraint follows.

We will now proceed to the quantization of the twisted torus. For the sake of generality, we will ignore these constraints, and come back to them in appendix \ref{sec:fixcom}, used in section \ref{sec:ZZ}.

%%%%%%%%%%%%%%%%%%%%%%%%%%%%%%%
\subsection{Commutation relations for the  twisted torus}\label{sec:quanttor}

With the solutions to the classical equations at hand, we can now initiate a canonical quantization of a string propagating in the twisted torus, to linear order in $H$. We thus promote the expansion coefficients in the twisted torus solutions, that we derived in the previous section, to operators and deduce their commutation relations using the results of section \ref{sec:quantgen}. We limit ourselves to this part of the quantization procedure, since, as we will see in section \ref{sec:nongeo}, it gives enough information to show non-commutativity in the non-geometric background. For the same reason, we restrict our analysis to the commutators of $Y^1$ and $Y^2$ and their derivatives.

The canonical commutation relations \eqref{XX}-\eqref{XP}, that are the starting point for our analysis, contain a $\delta$ function. Accordingly, most of the relations in this section have to be understood in the sense of distributions. Thus, we use the following representation of the $\delta$ function
\beq
\delta(x)= \frac{1}{2\pi} \sum_{n \in \mathbb{Z}} e^{-i n x} \ .\label{eq:iddelta}
\eeq
Moreover, for functions $f: x \mapsto f(x)$ with compact support, the derivative of $\delta$ is defined as
\beq
f(x) \del_x \delta(x)\equiv - \delta(x) \del_x f(x) \ .
\eeq
As a consequence, since $\sigma \in [0,2\pi]$, we find
\beq \label{eq:deldelta}
(\sigma'-\sigma) \ \del_{\sigma} \delta(\sigma-\sigma')
= \delta(\sigma-\sigma') \ .
\eeq
On a similar tone, we will have to make use of the following
\beq
(\sigma-\sigma')\ \delta(\sigma-\sigma') = 0 \ . \label{eq:xdx}
\eeq
Finally, for any function $u(\tau,\sigma)=\sum_{n \in \Z} u_n\ e^{-\i n \sigma_+} $, where $u_n$ are constant, one can show that
\bea
\sum_{k\neq 0} e^{-\i k (\sigma - \sigma')} (u(\tau,\sigma) - u(\tau,\sigma'))
&= - (u(\tau,\sigma) - u(\tau,\sigma')) \ .
\eea
This can be reformulated, in the sense of distributions (using \eqref{eq:iddelta}), as
\beq
\delta (\sigma - \sigma') (u(\tau,\sigma) - u(\tau,\sigma')) = 0 \ . \label{eq:uss}
\eeq
This equation is also valid for the above function if $\sigma_+$ is exchanged for $\sigma_-$.

%%%%%%%%%%%%%%%%%%

\paragraph{Zeroth order}
After stating these identities, let us commence the canonical quantization. At zeroth order in $H$, we are left with the free string solution, and will of course only recover the standard commutation relations for the zero modes and oscillators of the free string. It is, however, useful to recall how these relations are derived, since we will use the same procedure at first order in $H$. To obtain the commutators between the coefficients, one should consider combinations of the coordinates and their derivatives which isolate a Fourier series with just one set of coefficients (e.g.~$\alpha_{n}^{\mu}$). By commuting such useful combinations, we then easily obtain the commutators of these coefficients from the canonical commutation relations. For instance, (using the shorthand $\epsilon_i = \pm 1$ for left and right movers) we find that
\beq
[\del_{\sigma_{\epsilon_1}} Y_0^{\mu} (\tau, \sigma) , \del_{\sigma_{\epsilon_2}'} Y_0^{\nu} (\tau, \sigma')]= \frac{\i \pi}{4} (\epsilon_1+\epsilon_2) \eta^{\mu\nu} \del_{\sigma} \delta(\sigma-\sigma') \ , \label{dY0dY0eps}
\eeq
where all three zeroth order canonical commutation relations \eqref{X0X0}-\eqref{YdtY0}, together with their derivatives, have been used, as well as the even parity of the $\delta$-function. For the free string, we have the useful relation
\beq
2 \del_{\sigma_{\epsilon}} Y_0^{\mu} = 2 p^{\mu}_{\epsilon} + \sum_{n \neq 0}  \alpha_{n \epsilon}^{\mu}\ e^{-i n \sigma_{\epsilon}} \ ,\label{combiH0}
\eeq
where we again use a shorthand notation for left and right movers:
$\alpha_{n\ +1}^{\mu}=\widetilde{\alpha}_n^{\mu}, \alpha_{n\ -1}^{\mu}=\alpha_n^{\mu} $, etc. Inserting this into \eqref{dY0dY0eps}, using \eqref{eq:iddelta},  and identifying the various Fourier coefficients, one reads off the commutators between the coefficients of \eqref{combiH0}. With this result at hand, one should go back to the canonical commutators \eqref{X0X0}-\eqref{YdtY0} and verify that they are satisfied. This is automatic for \eqref{dtX0dtX0}, while the other two fix the remaining commutators. The result is the following non-zero commutators of the free string:
\bea \label{eq:coefcom0}
& [\widetilde{\alpha}_m^{\mu}, \widetilde{\alpha}_n^{\nu}] = [\alpha_m^{\mu}, \alpha_n^{\nu}] = m\ \delta_{m,-n}\ \eta^{\mu\nu}  \ \; \; \forall m , n \in \mathbb{Z}^*\ ,\\
& [y^{\mu}, p^{\nu}]=\frac{\i}{2} \eta^{\mu\nu}  \ .\nn
\eea
This last commutator is usually decomposed into left and right movers
\beq
 [y_{\epsilon_1}^{\mu}, p_{\epsilon_2}^{\nu}] = \delta_{\epsilon_1 , \epsilon_2}\ \frac{\i}{4} \eta^{\mu \nu} \ . \label{xpLR}
\eeq
This division identifies the winding $N$ as the momentum associated to $\tilde{y}$, just as $p$ is associated to $y$, and is motivated by studies of T-duality of the free string. The other commutators of $y^{\mu}_{\epsilon}$ with the free string modes are taken to vanish. We will use these commutators in the following.

\paragraph{First order}
At first order in $H$, we proceed analogously. We first identify the following useful periodic combination from the solution \eqref{solY0Y0}
\bea
\!\!\!\!\! \Pi_{\epsilon}^{\mu} \equiv \del_{\sigma_{\epsilon}} Y_H^{\mu} & + \frac{1}{2} \del_{\sigma_{\epsilon}} Y_{0\epsilon}^{\nu} \left( -{\theta^{\mu}}_{\nu\rho} Y_{0(-\epsilon)}^{\rho} + {\theta^{\mu}}_{\nu\rho} (y^{\rho}_{-\epsilon} - \sigma_{\epsilon} p^{\rho}_{-\epsilon}) - 2 {\lambda^{\mu}}_{\nu\rho} \sigma_{\epsilon} (p^{\rho}_{\epsilon}-p^{\rho}_{-\epsilon} ) \right) \label{Pimu1}\\
&  + \frac{1}{2} \del_{\sigma_{\epsilon}} Y_{0\epsilon}^{\rho} \left( -{\theta^{\mu}}_{\nu\rho} Y_{0(-\epsilon)}^{\nu} + {\theta^{\mu}}_{\nu\rho} (y^{\nu}_{-\epsilon} - \sigma_{\epsilon} p^{\nu}_{-\epsilon}) \right) \nn\\
& \!\!\!\!\!\!\!\!\!\!\!\!\!\!\!\!\!\!\!\!\!\!\!\!\! = \frac{p_H^{\mu}}{2} + {\lambda^{\mu}}_{23} \frac{N^3 \epsilon}{2}\ (y^2 - \pi N^2)  +  \sum_{n\neq 0} \frac{e^{-\i n \sigma_{\epsilon}}}{2} \left(\g^{\mu}_{n \epsilon} + {\theta^{\mu}}_{\nu\rho}  \frac{\i}{n}\ p^{(\nu}_{-\epsilon} \alpha_{n\epsilon}^{\rho)} + {\lambda^{\mu}}_{\nu\rho} N^{\rho} \epsilon \frac{\i}{n}\ \alpha_{n\epsilon}^{\nu} \right) \label{Pimu2}
\eea
where, as above, $\epsilon=\pm1$ denotes left and right movers, and we refer to the line below \eqref{eqH1} for the parameters ${\theta^{\mu}}_{\nu\rho}$ and ${\lambda^{\mu}}_{\nu\rho}$.\footnote{To obtain the expression \eqref{Pimu2}, we have used that $\forall \epsilon\ ,\ \tilde{y}^2=-\epsilon y^2 + 2 \epsilon y^2_{\epsilon}\ , \ N^3=\epsilon(p_{\epsilon}^3 - p_{-\epsilon}^3)$, and also the convenient parametrization that $\forall \mu\ , \  {\lambda^{\mu}}_{23} N^3 Y_{0\epsilon}^2 = {\lambda^{\mu}}_{\nu\rho} N^{\rho} Y_{0\epsilon}^{\nu}$.}
This is more involved than the zeroth order counterpart \eqref{combiH0}, but can be used in a similar way. The main difference is that, since the first order canonical commutators, like \eqref{XXH}, are sums of two terms, we can only determine certain linear combinations of operator commutators. Fortunately, this is enough for our purposes: we derive these commutator relations in order to analyse the properties of the non-geometric T-dual of the twisted torus, and the linear combinations we find suffice for this analysis.

Thus, consider the sum of commutators
\beq \label{eq:Picom}
[H\ \Pi_{\epsilon_1}^{\mu} (\tau, \sigma), \del_{\sigma_{\epsilon_2}'} Y_{0}^{\nu} (\tau, \sigma')] +  [\del_{\sigma_{\epsilon_1}} Y_{0}^{\mu} (\tau, \sigma), H\ \Pi_{\epsilon_2}^{\nu} (\tau, \sigma')]
\eeq
for $(\mu,\nu)=(1,2)$ or $(2,1)$. On the one hand, using the periodic expression \eqref{Pimu2} of $\Pi_{\epsilon_1}^{\mu}$, as well as \eqref{combiH0} for $\del_{\sigma_{\epsilon_2}} Y_{0}^{\nu}$, and the zeroth-order commutators of coefficients, we obtain a sum of commutators between the first order coefficients ($p_H^{\mu}, \gamma^{\mu}_{n \epsilon} $) with the zeroth order coefficients ($\alpha^{\mu}_n ,\ldots$).\footnote{Note that commutators of first order coefficients among themselves only appear at second order in $H$.} On the other hand, we can use the definition \eqref{Pimu1} of $\Pi_{\epsilon}^{\mu}$ to equate \eqref{eq:Picom} with a combination of known zeroth- and first-order (canonical) commutators, in particular \eqref{dsY1dsY2H}. This is where considering a sum of commutators is necessary, so that we can use $H$-order information, as in \eqref{XXH}. Matching the two expressions thus obtained , one can now deduce the value of the commutators of coefficients entering the former. More precisely,
both expressions are Fourier series in $\tau$ and in $\sigma$, so each coefficient of those series should be matched. To see this, we must use both \eqref{eq:iddelta} and \eqref{eq:deldelta}.

Through this procedure, which is technically rather involved, we deduce the following commutators among modes of the $Y^1$ and $Y^2$ coordinates of the twisted torus:
\bea
\forall\ \epsilon_1,\epsilon_2,\ \forall\ &m\neq0,\ \forall\ n,k \neq0, k+n\neq 0\ ,\nn\\
& [p_H^{1}, p^{2}_{\epsilon_2}] = [p_H^{2}, p^{1}_{\epsilon_1}] \ , \label{eq:comD0pe}\\
& [\g^1_{m \epsilon_1}, p^2_{\epsilon_2}] - \frac{1}{2} [p_H^2, \alpha^1_{m\epsilon_1}] = [\g^2_{m \epsilon_1}, p^1_{\epsilon_2}] - \frac{1}{2} [p_H^1, \alpha^2_{m\epsilon_1}] = \frac{\i}{8} \alpha^3_{m \epsilon_1} \ , \label{eq:comCpeD0a}\\
& [\g^1_{m \epsilon_1}, \alpha^2_{-m \epsilon_2}] - [\g^2_{-m \epsilon_2}, \alpha^1_{m \epsilon_1}] = \delta_{\epsilon_1 , \epsilon_2} \left(y^3 m -\frac{\i N^3 \epsilon_1}{2} \right) \ , \label{eq:comCamm}\\
& [\g^1_{k \epsilon_1}, \alpha^2_{n\epsilon_2}] - [\g^2_{n \epsilon_2}, \alpha^1_{k\epsilon_1}] = \frac{\i}{4} \frac{k-n}{k+n}\ \delta_{\epsilon_1,\epsilon_2}\ \alpha^3_{(k+n) \epsilon_1}\ .\label{eq:comCakn}
\eea
Using the freedom to choose $\epsilon_1$ and $\epsilon_2$, one can actually deduce from \eqref{eq:comD0pe} and \eqref{eq:comCpeD0a} the following $\forall\ \epsilon,\ \forall\ m\neq0$
\bea
& [p_H^{1}, N^{2}] = [p_H^{2}, N^{1}] = [p_H^{1}, p^{2}] - [p_H^{2}, p^{1}] = 0 \label{eq:comD0N}\\
& [\g^1_{m \epsilon}, N^2] = [\g^2_{m \epsilon}, N^1] = 0 \label{eq:comCN}\\
& [\g^1_{m \epsilon}, p^2] - [p_H^2, \alpha^1_{m\epsilon}] = [\g^2_{m \epsilon}, p^1] - [p_H^1, \alpha^2_{m\epsilon}] = \frac{\i}{4} \alpha^3_{m \epsilon} \ . \label{eq:comCpD0a}
\eea

As for the free string, we can record further conditions on the commutators between modes by systematically inserting the relations just derived into the canonical commutators. We start with commutators having more derivatives, and proceed to those with no derivative, i.e. we begin with \eqref{dtYdtYH}, continue to  \eqref{Y1dY2H}, and finally study \eqref{XXH}. Actually, using \eqref{eq:iddelta} and \eqref{eq:uss} and the commutators above, we find that \eqref{dtYdtYH} is already satisfied (just as at zeroth order), so the next set of conditions is obtained from \eqref{Y1dY2H}. Making use of \eqref{eq:xdx} and \eqref{eq:uss}, we find that this commutator implies that, $\forall \epsilon$, $ \forall n \neq 0$,
\bea
& [y^1, p_H^2] + [y_H^1, p^2] = [y^2, p_H^1] + [y_H^2, p^1] = \frac{\i}{2} y^3 \ , \label{eq:comyD}\\
& [y^1, \g_{n \epsilon}^2] + [y_H^1, \alpha_{n  \epsilon}^2] = [y^2, \g_{n \epsilon}^1] + [y_H^2, \alpha_{n  \epsilon}^1] = -\frac{1}{8n} \alpha_{n  \epsilon}^3 \ . \label{eq:comyCn}
\eea
As the last step of the analysis, we turn to \eqref{XXH}, i.e. $[Y^1(\sigma,\tau), Y^2(\sigma',\tau)]=0$. Given the relations above, this holds if we impose
\bea \label{eq:comzeromod}
&[y^1,y_H^2] - [y^2,y_H^1] = 0 \\
 &[N^1,y_H^2] = [N^2,y_H^1] = 0 \ .\nn
\eea

This ends the derivation of the $(1,2)$ or $(2,1)$ commutators. We have verified that the set of commutators derived on the coefficients, namely \eqref{eq:comD0pe} to \eqref{eq:comCpD0a}, together with \eqref{eq:comyD}, \eqref{eq:comyCn}, and \eqref{eq:comzeromod}, is equivalent to the ${\cal O}(H)$ canonical commutation relations \eqref{XXH}, \eqref{Y1dY2H} and \eqref{dtYdtYH}. This is very similar to the zeroth order result.

As a final remark, note that we find non-trivial commutators between modes of $Y^1$ and $Y^2$ at linear order in $H$. This is in contrast to the zeroth order commutators, which always vanish between operators associated to different directions. As we will see in the next section, these non-vanishing commutators are crucial for the non-commutativity of the non-geometric background.

\section{Analysis of the non-geometric background}\label{sec:nongeo}

Here, we turn to our main goal: the analysis of commutators between coordinates in a non-geometric background. We will make use of all results derived in previous sections. First, we use the relations of section \ref{sec:Tdual} to T-dualise the twisted torus along the $Y^2$ coordinate. This results in a non-geometric background, where we will be able to give the explicit mode expansion of the string coordinates $Z^\mu$. Second, leaving some details to the appendix \ref{sec:fixcom}, we work-out an expression for $[Z^1(\tau,\sigma),Z^2(\tau,\sigma')]$, which turns out to be nonzero in the limit $\sigma' \rightarrow \sigma$, and thus the coordinates fail to be commutative. We finally discuss the origin of this non-commutativity, and interpret it using in particular the non-geometric $Q$-flux.\\

Before getting started, let us recall a few things on this background. We have seen in section \ref{sec:nongeobg} that performing a T-duality transformation on the twisted torus leads to a field configuration (here in rescaled notation)
\beq\label{eq:nongeoconfigrescaled}
G = f \begin{pmatrix}  1 & 0 & 0 \\ 0 & 1 & 0 \\ 0 & 0 & f^{-1} \end{pmatrix}\ , \ \
B = f \begin{pmatrix} 0 & -HZ^3 & 0 \\ HZ^3 & 0 & 0 \\ 0 & 0 & 0 \end{pmatrix}\ , \ \
f=\left(1+\left(H Z^3 \right)^2 \right)^{-1}\ ,
\eeq
where we now denote the coordinates as $Z^{\mu}$, and still identify $Z^3_0 = Y^3_0$. Taking into account the factor $f$, it is easy to check that neither $G$ nor $B$ respect the periodicity of $Z^3$. To be precise, we cannot find an atlas of the above space such that the fields $G$ and $B$ are patched only by diffeomorphisms and gauge transformations. Including T-duality in the set of allowed transition functions amends this problem. So, this background is a non-geometric background and can be viewed as a T-fold \cite{Hull:2004in}. An alternative take on the problem is to perform a field redefinition, that replaces the ill-defined fields $G, B$ with globally defined objects \cite{Andriot:2011uh,Andriot:2012wx,Andriot:2012an}. One of those is the non-geometric $Q$-flux mentioned in the introduction.

Interestingly, to linear order in $H$, the above configuration \eqref{eq:nongeoconfigrescaled} is equivalent to the flat torus with $H$-flux (see table \ref{tab:rescandH}). It seems possible that the non-geometric properties are invisible at this order, but this naive expectation is wrong. In fact, the difference between the two situations is visible in the boundary conditions of the coordinates, as discussed in section \ref{sec:Tdual}. The classical solutions thus are different and so are the commutators between coordinates. This is consistent with the idea that non-geometry is related to global aspects, and therefore to the boundary conditions.

\subsection{Classical solution from T-duality}

We will now use the T-duality relations \eqref{eq:TdcoordYZ} to derive the coordinate expansions $Z^{\mu}$. These relations give explicit expressions for the derivatives of $Z^\mu$ in terms of the solutions for $Y^\mu$.  Integrating these with respect to $\tau$ or $\sigma$, we obtain expressions for $Z^\mu$. As an integrability check, we compute the derivatives of the relations \eqref{eq:TdcoordYZ}:
\begin{eqnarray}
\partial_{\sigma} \partial_{\tau} Z^2&=&\partial_{\sigma}\partial_{\sigma} Y^2-H\partial_{\sigma}(Y^3\partial_{\sigma} Y^1)\, ,\quad
\partial_{\tau} \partial_{\sigma} Z^2=\partial_{\tau} \partial_{\tau} Y^2-H\partial_{\tau} (Y^3\partial_{\tau} Y^1)\, .
\end{eqnarray}
Subtracting these two expressions, and keeping terms to linear order in $H$, we recover the equation of motion for $Y^2$ (the same phenomenon as in \eqref{eq:trivialeom}). We thus expect to find a consistent expression for $Z^2$ up to ${\cal O}(H^2)$ terms.

Our approach now allows to determine the $Z^\mu$ order by order in $H$,
\beq
Z^1 = Z^1_0 + H Z^1_H \ , \ Z^2 = Z^2_0 + H Z^2_H \ .
\eeq

\paragraph{Zeroth order}

At zeroth order, the T-duality relations \eqref{eq:TdcoordYZ} integrate to
\bea
Z^1_0(\tau,\sigma) & = z^1-y^1 + Y^1_0(\tau,\sigma) \\
Z^2_0(\tau,\sigma) & = z^2-\tilde{y}^2 + \tilde{Y}^2_0(\tau,\sigma)
\, ,\label{eq:Z20}
\eea
where $z^1, z^2$ are so far arbitrary integration constants. They will be related to $y^1, \tilde{y}^2$ on a quantum level in section \ref{sec:ZZ}. We conclude that, at zeroth order in $H$, T-duality essentially changes the sign of the right-moving part of $Y^2_0$, as expected. The functions $Z^1_0$ and $Z^2_0$ are again free string solutions. For completeness, let us recall that we take $Z^3_0=Y^3_0$.

\paragraph{First order}

At the first order in $H$, the T-duality relations \eqref{eq:TdcoordYZ} reduce to
\bea\label{eq:tdualityH}
& \partial_{\tau} Z^2_H = \partial_{\sigma} Y^2_H-HY^3_0\partial_{\sigma} Y^1_0\, ,\quad\partial_{\sigma} Z^2_H = \partial_{\tau} Y^2_H-HY^3_0\partial_{\tau} Y^1_0\, ,\\
& \partial_{\tau} Z^{1,3}_H = \partial_{\tau}  Y^{1,3}_H\, ,\quad\partial_{\sigma} Z^{1,3}_H = \partial_{\sigma}Y^{1,3}_H\, ,\nn
\eea
where $Y^{\mu}_0$ are the zeroth order solutions \eqref{eq:sol0} and $Y^\mu_H$ are given by \eqref{finalsol}. It is easy to obtain $Z^1_H$ as
\beq \label{eq:Z1H}
Z^1_H(\tau,\sigma)=z_H^1-y_H^1 + Y^1_H(\tau,\sigma) \ ,
\eeq
where $z_H^1$ is an arbitrary integration constant. Also $Z^2_H$ can be obtained from a straightforward computation, followed by a rearrangement of the solution,
\bea \label{eq:tX2Hnew}
Z^2_H(\tau,\sigma) =\; &z_H^2 + p_H^2 \sigma
+\frac{\i}{2}\sum_{n\neq0} \frac{1}{n} \Big(
\tg_n^{2} e^{-\i n\sigma_+}  - \g_n^{2}e^{-\i n\sigma_-}
\Big) \\ \nn
&-\frac{1}{4} \left( Y_0^{3}|_{\Sigma} \tilde{Y}_0^{1}|_{\Sigma}
-
Y_0^{1}|_{\Sigma} \tilde{Y}_0^{3}|_{\Sigma}
\right)
\\ \nn
&+\frac{1}{2} \tau \left(
p^1 \tilde{Y}_0^{3}|_{\Sigma} + p^3 \tilde{Y}_0^{1}|_{\Sigma}
-N^1 Y_0^{3}|_{\Sigma} - N^3 Y_0^{1}|_{\Sigma}
\right)-(y^3 + p^3\tau + N^3 \sigma)\tilde{Y}_0^{1}|_{\Sigma}\phantom{\frac{1}{1}}\\ \nn
&- p^1y^3\sigma
-N^1(y^3 + N^3 \sigma)\tau  -\frac{1}{2}( N^1 p^3 \tau^2 + p^1N^3 \sigma^2)
\\ \nn
&-\frac{\i}{4} \sum_{n\neq0} \frac{1}{n}
\left[\sigma_+ \tilde{\alpha}^1_{-n} \tilde{\alpha}^3_n -
\sigma_- \alpha^1_{-n} \alpha^3_n  \right]
 \\ \nn
&+\frac{1}{2}\sum_{n\neq0} \frac{1}{n^2} \Big(
\left[ p^1_L \tilde{\alpha}^3_n - p^3_L \tilde{\alpha}^1_n \right]e^{-\i n\sigma_+}
- \left[p^1_R \alpha^3_n - p^3_R \alpha^1_n \right]e^{-\i n\sigma_-}
\Big)  \\ \nn
&+\frac{1}{4}\sum_{\substack{n,m\neq0\\ m\neq -n}} \frac{1}{n(n+m)}
\left[\tilde{\alpha}^1_m \tilde{\alpha}^3_n e^{-\i (n+m)\sigma_+} -
\alpha^1_m \alpha^3_n e^{-\i (n+m)\sigma_-} \right]
\, .
\eea
Note that $z_H^2$ is an undetermined integration constant. From this expression, we obtain the boundary condition for $Z^2$
\bea\label{zztrans}
Z^2(\tau,\sigma+2\pi) =&Z^2(\tau,\sigma) + 2 \pi p^2 \\
+&H\Big( 2 \pi N^3 (\tilde{y}^1-\tilde{Y}^1_0) + 2\pi (p_H^2 - p^1y^3 - p^1N^3 \pi)
 -\frac{\i\pi}{2} \sum_{n\neq0} \frac{1}{n} \left[\tilde{\alpha}^1_{-n} \tilde{\alpha}^3_n + \alpha^1_{-n} \alpha^3_n  \right]\Big)\nn
 \, ,
\eea
where by $\tilde{Y}^1_0$ we mean the quantity in \eqref{eq:sol0LR}. One can verify that equation \eqref{zztrans} is in full agreement with the boundary conditions found in \eqref{eq:bdyZ monod} up to a constant shift, as expected. Finally, we note that $Z^2_H$ can be rewritten in the following form
\beq
Z^2_H = \frac{1}{4}  \left(Y^1_0 \tilde{Y}^3_0 - Y^3_0 \tilde{Y}^1_0 \right) + g_L(\sigma_+) + g_R(\sigma_-) \ ,\label{formZ2H}
\eeq
where $g_L$ and $g_R$ are functions of the left- or right-moving coordinate only. As discussed for \eqref{solY0Y0}, this rewriting makes manifest that $Z^2_H$ is a solution to the equations of motion.

\paragraph{Summary}

These are the coordinate solutions obtained after T-dualising on $Y^2$:
\bea\label{eq:Zsum}
Z^1 (\tau,\sigma) &= z^1-y^1 +Y^1_0(\tau,\sigma) + H (z_H^1 - y_H^1 + Y^1_H(\tau,\sigma)) \\ \nn
Z^2 (\tau,\sigma) &= z^2-\tilde{y}^2 + \tilde{Y}^2_0(\tau,\sigma) + H Z^2_H(\tau,\sigma) \\ \nn
Z^3 (\tau,\sigma) &= Y^3_0(\tau,\sigma) + H (z_H^3 - y_H^3 + Y^3_H(\tau,\sigma)) \, ,
\eea
where the zeroth order, free string, solutions are given by \eqref{eq:sol0LR}, $Y^{\mu}_H$ are given by \eqref{finalsol} or \eqref{solY0Y0}, $Z^2_H$ is given by \eqref{eq:tX2Hnew}, and $z^{\mu}, z_H^{\mu}$ are integration constants.

\subsection{Commutator of coordinates}\label{sec:ZZ}

We will now use the explicit solutions \eqref{eq:Zsum} to compute the coordinates commutator
\beq\label{eq:whatwewant}
[Z^1(\tau,\sigma), Z^2(\tau,\sigma')] \ .
\eeq
The result is presented in \eqref{eq:resultsimpl} and discussed in section \ref{sec:ZZdisc}, to which we direct readers that are not interested in the computional details. The general procedure is that we take over the commutators \eqref{eq:coefcom0}, \eqref{xpLR}, \eqref{eq:comD0pe} - \eqref{eq:comCpD0a} from the twisted torus background and use them here to conclude what the commutator of interest has to be. In some sense this is the reverse procedure compared to what was done in section \ref{sec:quanttor}. Recall that in table \ref{tab:rescandH} we find that for vanishing $H$, all three T-dual situations reduce to the free string on a flat torus. This is a well-understood, geometric setting where the canonical commutator relations hold. We can thus conclude that non-commutativity is only possible at linear or higher order in $H$.

Let us start computing the commutator \eqref{eq:whatwewant} by plugging in the explicit solutions \eqref{eq:Zsum}. We have
\bea
[Z^1, Z^2] & = [z^1- y^1, z^2-\tilde{y}^2] + [z^1- y^1, \tilde{Y}^2_0] + [Y^1_0, z^2-\tilde{y}^2] \label{eq:comfinalgen}\\
& \quad + H \Big( [Z^1_H, z^2-\tilde{y}^2] + [z^1- y^1, Z^2_H] \Big)  + H \Big( [Z^1_H, \tilde{Y}^2_0] + [Y^1_0, Z^2_H] \Big)\ , \nn
\eea
where we omitted $(\tau,\sigma)$ and $(\tau,\sigma')$ for simplicity.

Given the argument above, we require that non-commutativity must not stem from zeroth order commutators. Therefore, we set
\beq
[Z^1_0,Z^2_0]=0 \ .
\eeq
This restricts the undetermined commutators at zeroth order to
\beq
[z^1 - y^1,\; \text{any } 0^{\text{th}} \text{ order operator}] = [z^2 - \tilde{y}^2,\; \text{any } 0^{\text{th}} \text{ order operator}] = 0\ .
\eeq
In other words, we require that $z^1$ ($z^2$) has the same zeroth order commutators as $y^1$ ($\tilde{y}^2$). It follows, that the first line of \eqref{eq:comfinalgen} vanishes. The first term on the second line simplifies to
\bea \label{eq:simplifZZ}
[Z^1_H (\tau, \sigma), &z^2-\tilde{y}^2] + [z^1- y^1, Z^2_H (\tau, \sigma')] \\ 
&= \Big[z_H^1 + p_H^1 \ \tau +\frac{\i}{2}\sum_{n\neq0}\frac{1}{n} \left(\tg_n^1 e^{-\i n\sigma_+} + \g_n^1 e^{-\i n\sigma_-} \right) , z^2-\tilde{y}^2 \Big] \nn \\ 
&\quad+ \Big[z^1-y^1, z_H^2 + p_H^2\ \sigma' + \frac{\i}{2}\sum_{n\neq0} \frac{1}{n} \Big( \tg_n^{2} e^{-\i n\sigma_+'}  - \g_n^{2}e^{-\i n\sigma_-'} \Big)\Big] \ , \nn
\eea
using \eqref{eq:Z1H} and \eqref{eq:tX2Hnew}. The most involved computation is the last term of \eqref{eq:comfinalgen}, so we give it in two parts. Using \eqref{eq:tX2Hnew} and all the commutators discussed so far, we compute
\bea
[Y^1_0 (\tau, \sigma) , Z^2_H (\tau, \sigma')] & = \Big[Y^1_0 (\tau, \sigma), z_H^2 + p_H^2\ \sigma' + \frac{\i}{2}\sum_{n\neq0} \frac{1}{n} \Big( \tg_n^{2} e^{-\i n\sigma_+'}  - \g_n^{2}e^{-\i n\sigma_-'} \Big)\Big] \label{eq:Y10Z2Hb}\\
&\quad - \frac{\i}{4} N^3 \sum_{n\neq0} \frac{1}{n^2} e^{-\i n (\sigma' - \sigma)} \nn\\
&\quad +\frac{1}{8} \left(y^3 + N^3 (3 \sigma'-2\sigma) - p^3 \tau + Y^3_0(\sigma') +2 Y^3_0(\sigma) \right) \sum_{n\neq0} \frac{1}{n} e^{-\i n (\sigma' - \sigma)} \nn\\
&\quad + \frac{\i}{4} \sigma' \left(-y^3 + N^3 (\sigma - \sigma') - Y^3_0(\sigma) \right) + \frac{\i}{4} \tau \left(p^3 \sigma + \tilde{Y}^3_0(\sigma') - \tilde{Y}^3_0(\sigma) \right) \ .\nn
\eea
Similarly, using \eqref{eq:Z1H}, we compute
\bea
[Z^1_H (\tau, \sigma) , \tilde{Y}^2_0 (\tau, \sigma')] & = \Big[z_H^1 + p_H^1 \ \tau +\frac{\i}{2}\sum_{n\neq0}\frac{1}{n} \left(\tg_n^1 e^{-\i n\sigma_+} + \g_n^1 e^{-\i n\sigma_-} \right) , \tilde{Y}^2_0 (\tau, \sigma')\Big] \label{eq:Z1HtY02}\\
&\quad -\frac{1}{8} \left(-y^3 + 3 N^3 \sigma + p^3 \tau + Y^3_0(\sigma) \right) \sum_{n\neq0} \frac{1}{n} e^{-\i n (\sigma' - \sigma)} \nn\\
&\quad + \frac{\i}{4} \sigma \left( 2\pi N^3 - p^3 \tau \right) + \frac{\i}{4} \tau \left( \tilde{Y}^3_0(\sigma) -\tilde{y}^3 \right)  - \frac{\i}{4} N^3 (\tau^2 + \sigma^2 - 2 \sigma \sigma') \
 .\nn
\eea
The sum of the first lines of \eqref{eq:Y10Z2Hb} and \eqref{eq:Z1HtY02} exactly contain the right combinations of commutators to use \eqref{eq:comD0pe}, and \eqref{eq:comCamm} to \eqref{eq:comCpD0a}. Performing that replacement and compiling all results, we eventually obtain for the commutator of interest
\bea
 \frac{1}{H} [Z^1 (\tau, \sigma) , Z^2 (\tau, \sigma')]  \label{eq:comfinalb}
& = \Big[z^1 , z_H^2\Big] + \Big[z_H^1 , z^2 \Big] 
+ \tau \Bigg( \Big[ p^1 , z_H^2\Big]  + \Big[z_H^1 , N^2 \Big] + \Big[p_H^1, z^2\Big] \Bigg)\\
&\quad + \sigma \Bigg( \Big[ N^1 , z_H^2\Big]+ \frac{\i \pi}{2} N^3 \Bigg)  - \sigma' \Bigg( \Big[p^2, z_H^1 \Big] + \Big[p_H^2, z^1 \Big] + \frac{\i}{2} y^3 \Bigg) \nn\\
&\quad + \frac{\i}{2}\sum_{n\neq0}\frac{1}{n} \Bigg(e^{-\i n\sigma_+} \left( \Big[\widetilde{\alpha}_n^{1}, z_H^2\Big] + \Big[ \tg_n^1 , z^2\Big] \right) + e^{-\i n\sigma_+'} \left( \Big[z_H^1 , \widetilde{\alpha}_n^{2} \Big] + \Big[z^1,  \tg_n^{2} \Big] \right) \Bigg) \nn\\
&\quad + \frac{\i}{2}\sum_{n\neq0}\frac{1}{n} \Bigg(e^{-\i n\sigma_-} \left( \Big[ \alpha_n^{1} , z_H^2\Big] + \Big[ \g_n^1  , z^2\Big] \right) - e^{-\i n\sigma_-'} \left( \Big[z_H^1 , \alpha_n^{2} \Big] + \Big[z^1,  \g_n^{2} \Big] \right) \Bigg) \nn\\
&\quad + \frac{\i}{16} \sum_{n\neq0} \frac{1}{n^2} \left( \widetilde{\alpha}^3_n (e^{-\i n\sigma_+'}-e^{-\i n\sigma_+}) - \alpha^3_n (e^{-\i n\sigma_-'}-e^{-\i n\sigma_-}) \right) \nn\\
&\quad - \frac{\i}{2} N^3 \sum_{n\neq0} \frac{1}{n^2} e^{-\i n (\sigma' - \sigma)} + \frac{1}{2} N^3 (\sigma' -\sigma) \sum_{n\neq0} \frac{1}{n} e^{-\i n (\sigma' - \sigma)} - \frac{\i}{4} N^3 (\sigma' - \sigma)^2 \ .\nn
\eea

Before we proceed, a few comments are in order.
\begin{itemize}
\item From the T-duality relations \eqref{eq:TdcoordYZ} and using \eqref{YdtY0}, \eqref{dsY1dtY2H} we can immediately deduce
\beq
[\del_{\sigma} Z^1(\tau,\sigma), \del_{\sigma'} Z^2(\tau,\sigma')] = [\del_{\sigma} Y^1(\tau,\sigma), (\partial_{\tau} Y^2 - HY^3\partial_{\tau} Y^1)(\tau,\sigma')] = 0 \ , \label{dsZ1dsZ2}\\
\eeq
without referring to any particular mode expansion. This has to be understood in the sense of distributions and implies that the commutator between $Z^1$ and $Z^2$ can be written, up to possible contributions from distributions, as
\beq
[Z^1(\tau,\sigma), Z^2(\tau,\sigma')] = f_1(\tau, \sigma) + f_2(\tau, \sigma') \ ,\label{Z1Z2}
\eeq
where $f_1, f_2$ are arbitrary functions from this perspective. In other words, the $\sigma$- and $\sigma'$-dependence of this commutator has to be separable. This holds for the expression \eqref{eq:comfinalb}, up to the last line, but one can verify by using \eqref{eq:iddelta} and \eqref{eq:deldelta} that
\bea
& \del_{\sigma} \del_{\sigma'}\left(- \frac{\i}{2} N^3 \sum_{n\neq0} \frac{1}{n^2} e^{-\i n (\sigma' - \sigma)} + \frac{1}{2} N^3 (\sigma' -\sigma) \sum_{n\neq0} \frac{1}{n} e^{-\i n (\sigma' - \sigma)} - \frac{\i}{4} N^3 (\sigma' - \sigma)^2 \right) \nn \\
& = \i \pi N^3 \left(\delta(\sigma' - \sigma) + (\sigma' - \sigma) \del_{\sigma'} \delta(\sigma' - \sigma) \right) \nn \\ & 
= 0 \ ,\label{deldelmagic}
\eea
to be understood in the sense of distributions. We conclude that \eqref{dsZ1dsZ2} is fulfilled, which is a non-trivial check of our result \eqref{eq:comfinalb}. This comment also indicates that the last line of \eqref{eq:comfinalb} is very special.

\item All the remaining commutators in \eqref{eq:comfinalb} are undetermined by our construction. The reason is that they involve the new integration constants $z^{1,2}$ and $z_H^{1,2}$ for which we do not have information in the T-dual backgrounds. In appendix \ref{sec:fixcom}, we fix the value of these unknown commutators by using some physical arguments and reasonable analogies with situations we know. Nevertheless, in absence of a more fundamental guideline, this fixing can strictly speaking be considered as a restriction or a subcase among other possibilities. At least, the result obtained makes this subcase interesting to consider: the particular choice of the undetermined commutators we argue for leads to non-commuting coordinates.
\end{itemize}
Then, using the values given in \eqref{eq:N2z1H} - \eqref{eq:a2z1H} and \eqref{eq:z1z2H+} - \eqref{eq:a1z2H} for the remaining commutators, we eventually reduce \eqref{eq:comfinalb} to
\bea
& \frac{1}{H} [Z^1 (\tau,\sigma) , Z^2 (\tau,\sigma')] \label{eq:resultsimpl} \\
& = - \frac{\i}{2} N^3 \sum_{n\neq0} \frac{1}{n^2} e^{-\i n (\sigma' - \sigma)} + \frac{1}{2} N^3 (\sigma' -\sigma) \sum_{n\neq0} \frac{1}{n} e^{-\i n (\sigma' - \sigma)} - \frac{\i}{4} N^3 (\sigma' - \sigma)^2  \ .\nn
\eea
This expression is the one we consider from now on.

\subsection{Non-commutativity}\label{sec:ZZdisc}

In the previous section, we computed the commutator between $Z^1$ and $Z^2$, and finally got to the result given in \eqref{eq:resultsimpl}. From this, one can easily infer the appearance of non-commutativity in the limit $\sigma' \rightarrow \sigma$
\beq
[Z^1 (\tau,\sigma) , Z^2 (\tau,\sigma')] \xrightarrow{\sigma' \rightarrow \sigma}  - \frac{\i}{2} \frac{\pi^2}{3} N^3 H \ .\label{eq:limNC}
\eeq
We will discuss this result in terms of non-geometric fluxes and the conjecture \eqref{commcloseda}, and mention an associated effective action. But before that, let us spell out the precise origin of the non-commutativity, and show why it does not occur in the other two geometric backgrounds.

\subsubsection{Origin of non-commutativity}\label{sec:origin}

As can be seen from \eqref{eq:resultsimpl}, the only relevant part of the commutator $[Z^1(\tau,\sigma), Z^2(\tau,\sigma')]$ is given by
\beq\label{eq:piece}
A \equiv -\frac{\i}{2}N^3\sum_{n\neq0}\frac{1}{n^2}e^{-\i n(\sigma'-\sigma)}\ .
\eeq
In this section we give a guideline to the details of how $A$ arises in our analysis, which shows that T-duality plays a dominant role in this explanation. In what follows, we first spot the origin of $A$ in the non-geometric situation by tracing two different contributions. After that, we show how T-duality induces subtle changes such that there is no $A$ in the two geometric backgrounds.

\paragraph{Non-geometric background}

There are two different contributions to $A$, each of them adding one half of it.
\begin{enumerate}[a)]
\item The first contribution can be seen in the second line of $[Y^1_0,Z^2_H]$, \eqref{eq:Y10Z2Hb}. It comes from the zeroth order commutator,
\beq [\alpha^1_{m\epsilon},\alpha^1_{n\epsilon}] = m\ \delta_{m,-n}\ ,\label{aacomoncemore}
\eeq
where one of the $\alpha^1_m$ comes from $Y_0^1$, and the other can be traced back to a particular piece in $Z^2_H$, namely the sixth line of \eqref{eq:tX2Hnew},
\beq
+\frac{1}{2}\sum_{n\neq0} \frac{1}{n^2} \Big(
\left[ p^1_L \tilde{\alpha}^3_n - p^3_L \tilde{\alpha}^1_n \right]e^{-\i n\sigma_+}
- \left[p^1_R \alpha^3_n - p^3_R \alpha^1_n \right]e^{-\i n\sigma_-}
\Big)\ .
\eeq
After using \eqref{aacomoncemore}, one is left with two terms that add up to a piece proportional to $-(p^3_L-p^3_R)=-N^3$, giving $\tfrac{1}{2}A$.

This contribution appears as a particular feature of T-duality in the following sense: the above term can be characterized by noting its $1/n^2$ dependence. Such a dependence originates from the third line of the solution in the twisted torus frame, \eqref{finalsol},
\beq
+{\theta^{\mu}}_{\nu\rho}\ \frac{1}{2}\ \tau\ \left(p^{\rho} Y_0^{\nu}|_{\Sigma} - N^{\rho} \tilde{Y}_0^{\nu}|_{\Sigma} + p^{\nu} Y_0^{\rho}|_{\Sigma} - N^{\nu} \tilde{Y}_0^{\rho}|_{\Sigma} \right)\ ,\label{thirdline}
\eeq
and the particular form of the T-duality rules \eqref{eq:TdcoordYZ}. To be precise, the crucial point is the relation of $\sigma$-derivatives on $Z^2_H$ to $\tau$-derivatives on $Y^2_H$, and vice versa,
\beq
\partial_\tau Z^2_H = \partial_\sigma Y^2_H + \dots\ ,\quad \partial_\sigma Z^2_H = \partial_\tau Y^2_H + \dots \ ,
\eeq
that after integration produces from \eqref{thirdline}, amongst others, terms with a $1/n^2$ dependence.
\item The second contribution comes from the commutator \eqref{eq:comCamm} between first order oscillators $\gamma_n^\mu$ and zeroth order oscillators $\alpha_n^\mu$,
\beq
[\g^1_{m \epsilon_1}, \alpha^2_{-m \epsilon_2}] - [\g^2_{-m \epsilon_2}, \alpha^1_{m \epsilon_1}] = \delta_{\epsilon_1 , \epsilon_2} \left(y^3 m -\frac{\i N^3 \epsilon_1}{2} \right) \ .\label{eq:gaagain}
\eeq
Applying it to one part of the first lines of \eqref{eq:Y10Z2Hb} and \eqref{eq:Z1HtY02},\\
\bea
&\Big[Y^1_0 (\tau,\sigma), \frac{\i}{2}\sum_{n\neq0} \frac{1}{n} \Big( \tg_n^{2} e^{-\i n\sigma_+'}  - \g_n^{2}e^{-\i n\sigma_-'} \Big)\Big] %\\
+ %&
\Big[\frac{\i}{2}\sum_{n\neq0}\frac{1}{n} \left(\tg_n^1 e^{-\i n\sigma_+} + \g_n^1 e^{-\i n\sigma_-} \right) , \tilde{Y}^2_0 (\tau,\sigma')\Big] \ , \nn
\eea
produces pieces that combine into $\tfrac{1}{2}A$. It has to be emphasised that due to the $\delta_{\epsilon_1,\epsilon_2}$ in \eqref{eq:gaagain} only commutators with either two right-moving or two left-moving oscillators are nonzero. As also several combinations of $\gamma_n^\mu$, $\tilde{\gamma}_n^\mu$ with $\alpha_n^\mu$, $\tilde{\alpha}_n^\mu$ appear, the result is very sensitive to the signs they come with. Schematically, we have
\bea \label{eq:osscheme}
[\tilde{\alpha}^1+\alpha^1,& \tilde{\gamma}^2-\gamma^2] + [\tilde{\gamma}^1+\gamma^1, \tilde{\alpha}^2-\alpha^2] \\
&= \left([\gamma^2,\alpha^1]-[\gamma^1,\alpha^2]\right) - \left([\tilde{\gamma}^2,\tilde{\alpha}^1]-[\tilde{\gamma}^1,\tilde{\alpha}^2]\right) \nn
 \\
&\quad + \left([\gamma^1,\tilde{\alpha}^2]-[\tilde{\gamma}^2,\alpha^1]\right) + \left([\gamma^2,\tilde{\alpha}^1]-[\tilde{\gamma}^1,\alpha^2]\right)\ .\nn
\eea
These are exactly the combinations we have at hand. Here, the two possible permutations for $\epsilon_1 = \epsilon_2$ add up, while the terms in the last row are simply zero.
\end{enumerate}
In table \ref{tab:contributions} we overview the fate of the two contributions a) and b), given in different lines, in the various T-dual backgrounds. The rightmost column depicts the discussion for the non-geometric background we have given so far, where the expression $1+1$ for the contribution b) pays tribute to the subtle sign combination explained above.

{\renewcommand{\arraystretch}{2}
\begin{table}[H]
\begin{center}
\begin{tabular}{|r||c|c|c|}
\hline
{\it Contribution} & \hspace{0.5cm}$H$-flux\hspace{0.5cm} & Twisted torus & Non-geometric \\
\hline\hline
a) $[\alpha,\alpha]$ & $-\frac{1}{2}$ & 0 & $\frac{1}{2}$ \\
b) $[\alpha,\gamma]$ & $\frac{1 + 1}{4}$ & $\frac{1 - 1}{4}$ & $\frac{1+1}{4}$ \\
\hline
{\bf Sum} & {\bf 0} & {\bf 0} & {\bf 1} \\
\hline
\end{tabular}
\caption{Contributions to \eqref{eq:piece} in units of $A$.}\label{tab:contributions}
\end{center}
\end{table}}
\vspace{-0.2in}

\subsubsection*{Geometric backgrounds}

For the twisted torus, two things change when recapitulating the above explanations. First, there is no type a) contribution - depicted by 0 in table \ref{tab:contributions}. This is most easily seen by noting the absence of any term with $N^3/n^2$ dependence in the expression for $Y_H^\mu$, \eqref{finalsol}, that could contribute. Second, the contribution b) is zero due to a sign change. As explained above, in the non-geometric situation there are two pieces coming from only left-moving and only right-moving oscillators. Here, they appear with the opposite sign and cancel out - depicted by $1-1$ in the table \ref{tab:contributions}. Schematically, this can be seen from
\bea
[\tilde{\alpha}^1+\alpha^1,& \tilde{\gamma}^2+\gamma^2] + [\tilde{\gamma}^1+\gamma^1, \tilde{\alpha}^2+\alpha^2] \\
&= -\left([\gamma^2,\alpha^1]-[\gamma^1,\alpha^2]\right) - \left([\tilde{\gamma}^2,\tilde{\alpha}^1]-[\tilde{\gamma}^1,\tilde{\alpha}^2]\right) \nn \\
&\quad + \left([\gamma^1,\tilde{\alpha}^2]-[\tilde{\gamma}^2,\alpha^1]\right) - \left([\gamma^2,\tilde{\alpha}^1]-[\tilde{\gamma}^1,\alpha^2]\right)\ ,\nn
\eea
where again the last row vanishes and now the opposite sign of the first term on the left-hand side causes the above mentioned cancellation.
The sign change exactly is the well-known sign change of the right-moving oscillators due to T-duality. In summary, there is no term \eqref{eq:piece} appearing in the twisted torus frame thanks to subtle adjustments from T-duality, which exactly meets our expectations.

For the torus with $H$-flux, both contributions appear, cf. the matching \eqref{eq:fixg1Xm+} and \eqref{eq:fixg1Xm-},
\beq \label{eq:matchgg}
\tg_{Xn}^1 = \tg^1_n + \frac{\i}{n} (p^3_L \widetilde{\alpha}^2_n -p^2_L \widetilde{\alpha}^3_n) + \dots\ , \quad \g_{Xn}^1 = -\g^1_n - \frac{\i}{n} (p^3_R \alpha^2_n -p^2_R {\alpha}^3_n) + \dots \ .
\eeq
The type a) pieces can be identified from some of the $1/n$ dependent terms in \eqref{eq:matchgg}. Commuting these with $X^2_0|_\Sigma$ produces a term proportional to $(p^3_L-p^3_R)=N^3$, similarly to the non-geometric situation, giving here $-\tfrac{1}{2}A$. For the type b) pieces, a similar combination of signs as in \eqref{eq:osscheme} leads to two parts adding up,
\bea
[\tilde{\alpha}^1-\alpha^1,& \tilde{\gamma}^2+\gamma^2] + [\tilde{\gamma}^1-\gamma^1, \tilde{\alpha}^2+\alpha^2] \\
&= \left([\gamma^2,\alpha^1]-[\gamma^1,\alpha^2]\right) - \left([\tilde{\gamma}^2,\tilde{\alpha}^1]-[\tilde{\gamma}^1,\tilde{\alpha}^2]\right) \nn \\
&\quad - \left([\gamma^1,\tilde{\alpha}^2]-[\tilde{\gamma}^2,\alpha^1]\right) - \left([\gamma^2,\tilde{\alpha}^1]-[\tilde{\gamma}^1,\alpha^2]\right)\ .\nn
\eea
Nevertheless, in total the two different contributions a) and b) appear with opposite signs and cancel out, as depicted in table \ref{tab:contributions}. Again, there is no term \eqref{eq:piece} remaining thanks to a rearrangement of signs from T-duality - as expected.

\subsubsection{Interpretation in terms of non-geometric fluxes and effective action}\label{sec:Qeffact}

In \cite{Andriot:2012an}, we conjectured that the non-commutativity should be given by
\beq\label{eq:conjecture}
[ Z^{\mu}(\tau,\sigma), Z^{\nu}(\tau,\sigma') ]_{\rm closed} \xrightarrow{\sigma' \rightarrow \sigma} \ c \ \i \oint_{C^{\rho}} Q_{\rho}{}^{\mu\nu} (Z)\ \d Z^{\rho} \ ,
\eeq
where $c$ is a numerical constant,\footnote{To be precise, the numerical factors and the $\i$ were missing in \cite{Andriot:2012an}.} and $C^{\rho}$ is a cycle, around which the closed string is wrapped $N^{\rho}$ times. There are some physical arguments in favour of this conjecture, and it would be nice to interpret the result of this paper \eqref{eq:limNC} in the same manner. Let us repeat it here for convenience
\beq
[Z^1 (\tau,\sigma) , Z^2 (\tau,\sigma')] \xrightarrow{\sigma' \rightarrow \sigma}  - \frac{\i}{2} \frac{\pi^2}{3} N^3 H \ .\label{eq:Z1Z2result}
\eeq
To allow the comparison between the two, we need, to start with, the $Q$-flux involved in \eqref{eq:conjecture}, so let us first recall how it is derived.

In \cite{Andriot:2011uh,Andriot:2012wx,Andriot:2012an}, we have proposed an effective action for (the NSNS sector of) non-geometric backgrounds, given in terms of a metric $\tilde{G}_{\mu\nu}$, an antisymmetric bivector $\beta^{\mu\nu}$ and a dilaton $\tilde{\phi}$. A nice feature of this action is that it involves the non-geometric $Q$- and $R$-fluxes, and in that way provides a lift of some gauged supergravities that previously did not have a higher dimensional origin. The $Q$-flux in particular is given by
\beq\label{Q}
Q_{\rho}{}^{\mu\nu} = \partial_{\rho}\beta^{\mu\nu}\ ,
\eeq
and in the case the background satisfies the condition $\beta^{\mu\nu}\partial_{\nu} =0$ when acting on arbitrary fields, the effective action then takes the form \cite{Andriot:2011uh}
\beq\label{Qaction}
\int \d x^d \sqrt{|\tilde{g}|} e^{-2\tilde{\phi}} \left(\tilde{\cal R} + 4 (\partial\tilde{\phi})^2 -\frac{1}{4} Q^2\right)\ ,
\eeq
where $(\partial\tilde{\phi})^2$ and $Q^2$ are simply the squares contracted with $\tilde{G}$. This action has the same form as the standard NSNS action. There is actually more to it: the effective action of \cite{Andriot:2011uh,Andriot:2012wx,Andriot:2012an} is equal (off-shell) to the NSNS action up to a total derivative term, provided one performs a field redefinition to go from one set of fields to the other. For instance, (the inverse of) $\tilde{G}$, and $\beta$, can be derived from the standard metric $G$ and $B$-field by considering the symmetric and antisymmetric parts of the following quantity
\beq\label{fieldredef}
(G+B)^{-1} = \tilde{G}^{-1} + \beta \ .
 \eeq
By studying global properties, we found that given a non-geometric background, the action we introduced can be better suited to describe the effective physics than the standard NSNS one. This statement works very well for the non-geometric background we consider in this paper, that was also studied in the appendix B of \cite{Andriot:2011uh}.\footnote{In particular, the condition $\beta^{\mu\nu}\partial_{\nu} =0$ holds for this background, so it admits the effective action \eqref{Qaction}.} The standard NSNS fields for the non-geometric background were given in \eqref{eq:nongeofields}. Using the field redefinition \eqref{fieldredef}, it is straightforward to compute
\beq \label{eq:tfields}
\tilde G_Q =   \begin{pmatrix} R_1^{-2} & 0 & 0 \\ 0 & R_2^{-2} & 0 \\ 0 & 0 & R_3^2 \end{pmatrix}
\ ,
\quad
\beta_Q =  \begin{pmatrix}  0 & H Z^3 & 0 \\ -H Z^3 & 0 & 0 \\ 0 & 0 & 0 \end{pmatrix}
\ ,
\eeq
from which one gets with \eqref{Q} the only non-trivial components\footnote{The $Q$-flux with flat indices is given here by $Q_3{}_{{\rm (flat)}}^{12} = H/(R_1R_2R_3)$; as mentioned already for the two geometric backgrounds, this is the quantity entering the T-duality chain \eqref{eq:TdualityChain} with respect to which we approximate. It is nice to verify that this definition of $Q$ is consistent with that claim. In addition, the conjecture equality \eqref{eq:conjecture} is invariant under the rescaling of table \ref{tab:resc} in appendix \ref{ap:not}, so we can use directly \eqref{eq:QH}.}
\beq\label{eq:QH}
Q_3{}^{12} = -Q_3{}^{21} = H  \; .
\eeq

We now have a $Q$-flux at hand for the non-geometric background, and we can compare our result to the conjecture \eqref{eq:conjecture}. We have a constant flux and the base cycle $C^3$ is just a circle wound $N^3$ times by the string. Thus, using \eqref{eq:QH},
\beq
[ Z^1(\tau,\sigma), Z^2(\tau,\sigma') ]_{\rm closed} \xrightarrow{\sigma' \rightarrow \sigma} \ c \ \i \ \oint_{C^3} Q_{3}{}^{12} \ \d Z^3 = 2\pi \ c \ \i\ H\ N^3 \ .\label{eq:conjecture2}
\eeq
This is in perfect agreement with our result \eqref{eq:Z1Z2result}, if we adjust accordingly the numerical factors: $c\ \i =-\frac{\pi}{6} \frac{\i}{2}$.

It is physically expected that the flux, as well as the winding, contribute to the RHS of the commutator, as we can see in the integrand of \eqref{eq:conjecture2}. Indeed, setting the flux to zero brings all the backgrounds to mere flat tori, for which the string is free and should clearly be commutative. Another way to argue is that the flux is responsible for the non-geometry in the last background, and so should have a non-trivial effect on the commutator. The winding should also be present: the non-trivial monodromies of the backgrounds, in particular the non-geometric one, only appears when going around the base circle, and for the string to probe this, it must extend in this direction. In other words, its winding should be non-zero to see a non-trivial effect. Finally, the numerical factors adjusted below \eqref{eq:conjecture2} should be mostly understood as coming from our conventions. For instance, all the non-trivial commutators derived from the canonical commutation relations in 
section \ref{sec:quantgen} come with an $\pi$ factor, so it is reasonable to get one as well. The $\frac{\i}{2}$ can also be understood as coming from our free string conventions as can be seen in \eqref{eq:coefcom0}.\\

\section{Conclusion}

The main result of this paper is that, for a closed string on a non-geometric background with $Q$-flux, some of its target space coordinates are non-commutative. Indeed, we have obtained
\bea
 [Z^1 (\tau,\sigma) , Z^2 (\tau,\sigma')] % \\
& = - \frac{\i}{2} N^3 H \left( \sum_{n\neq0} \frac{1}{n^2} e^{-\i n (\sigma' - \sigma)} + \i (\sigma' -\sigma) \sum_{n\neq0} \frac{1}{n} e^{-\i n (\sigma' - \sigma)} + \frac{1}{2} (\sigma' - \sigma)^2 \right) \nn\\
& \xrightarrow{\sigma' \rightarrow \sigma}  - \frac{\i}{2} \frac{\pi^2}{3} N^3 H \ .\label{eq:resultccl}%\nn
\eea
Our results confirm the first examples of closed string non-commutativity in relation with T-duality and non-geometry, which were found in \cite{Lust:2010iy,Condeescu:2012sp} by analysing
non-geometric string backgrounds with elliptic $Z_4$ monodromy. The class of backgrounds treated here are given by three-dimensional fibrations with parabolic monodromy transformations, when transporting a two-dimensional fibre along a base circle.

To discuss quantum properties of a closed string, such as its coordinate commutators, one has to choose a quantization method. For a geometric background, a standard option is canonical quantization (see e.g. \cite{quantrev} for recent reviews on other approaches \cite{Axenides:2008rn}). This amounts at first to impose the canonical commutation relations, which imply that the coordinates commute. On the contrary, for a non-geometric background, there is no reason to have the same relations. As discussed in the introduction, the literature even suggests that the coordinates do not commute. Therefore, to derive \eqref{eq:resultccl}, we instead used an indirect method. We considered a specific non-geometric background, related via T-duality to geometric situations where the canonical quantization tools could be used. More concretely, we started by solving the classical string on a three-dimensional torus with $H$-flux and on a T-dual twisted torus. We then performed the first steps of a canonical 
quantization on the latter, which 
resulted in the commutators of the various modes of the string (as is usually done for the free string). We finally used the T-duality relations between the twisted torus and the non-geometric background, to obtain the classical string on the latter in terms of the former. The commutators derived could then be used to compute $[Z^1, Z^2]$ as in \eqref{eq:resultccl}.

Let us make several remarks on this procedure. First, for these three T-dual spaces with flux to be consistent string backgrounds, one has to use the dilute flux approximation \eqref{eq:approx}. We therefore considered expansions of all quantities, in particular the world-sheet coordinates and the target space fields, to linear order in the flux parameter $H$. We showed that this parameter could be identified, according to the background, with respectively the $H$-flux, the geometric flux $f$, or the $Q$-flux (as given in the T-duality chain \eqref{eq:TdualityChain}). This method, leading to classical expressions for closed strings in non-trivial backgrounds (for instance the non-flat twisted torus) and associated quantum properties, is expected to be of more general use. As the one-loop $\beta$-functions of our sigma-models vanish, it would be interesting to derive similar results and properties using CFT techniques.

A second remark is that the T-duality relations \eqref{Tdcoordrel} between coordinates ${\cal X}$ and their duals $\hat{\cal X}$ played a crucial role in our analysis. We derived them following the Buscher procedure, but they were already considered in \cite{Duff:1989tf} and used since then in several different contexts. They could definitely serve other purposes. For instance, using those, one may get a more systematic definition of the doubled geometry (see \cite{Schulz:2011ye} and references therein for a discussion on that topic). It would be interesting to study this idea in more detail in the future.

The T-duality relations \eqref{Tdcoordrel} essentially map derivatives of coordinates. An important implication of this is that from T-duality, one can only deduce properties of the derivatives of coordinates: as we did, for instance, for the equations of motion, and the derivatives of commutators, in sections \ref{sec:Tdual} and \ref{sec:quantgen}. Going to the coordinate itself involves an integration, which can bring in important new information. For example, we discussed the new integration constants which can appear, and also some non-trivial distribution contributions. The last point is particularly relevant for the commutator of interest here: we showed in \eqref{dsZ1dsZ2} and below that $[\del_{\sigma} Z^1, \del_{\sigma'} Z^2]$ vanishes in the sense distributions, which makes the expression for its integrated counterpart \eqref{eq:resultccl} even more special. This discussion puts the common claim that T-duality is a canonical transformation in a new light. This claim would make it 
surprising, at a first glance, that a commutator of coordinates is not preserved under T-duality. However, since this transformation only relates derivatives of coordinates, such claims are subtle. As explained above, integration can bring important contributions and we should expect at most $[\del_{\sigma} Z^1, \del_{\sigma'} Z^2]$, and not $[Z^1, Z^2]$, to be preserved under T-duality, which is in excellent agreement with our results.\footnote{It is also worth noting that, in one reference on this topic, namely \cite{Alvarez:1994wj}, no $B$-field was considered, while for us the $B$-field plays a crucial role.}\\

The point of view of the doubled geometry \cite{Dabholkar:2005ve} offers another take on this discussion. In appendix \ref{sec:monod}, we argue on how to relate the target space monodromies to the boundary conditions of doubled closed string coordinates. Comparing the explicit expressions for the latter in the different backgrounds, namely \eqref{eq:bdyXmonod}, \eqref{eq:bdyYmonod} and \eqref{eq:bdyZ monod}, it is clear that a T-duality along $\iota$ exchanges ${\cal X}^{\iota}$ with $\hat{{\cal X}}^{\iota}$. Pushing this idea, one could consider a doubled phase space, with commutators among the standard coordinates and among the dual ones, and the T-duality would then exchange the two. If at first coordinates commute while the dual do not, this situation will be changed after T-duality. The non-commutativity then only ``arises'' because we focused on the subspace of standard coordinates; from the doubled space point of view, nothing really changes.

This is a way to understand what is happening here. We point out below \eqref{eq:bdyZ monod} that mixing standard and dual coordinates within boundary conditions is a sign of non-geometry, possibly leading to non-commutativity. This mixing occurs in the non-geometric background, and one can also see such an entanglement for the torus with $H$-flux, in the dual coordinates. Following this line of thoughts, we can conclude that the dual coordinates do not commute in this first geometric background, while the standard ones do, and this situation gets exchanged in the non-geometric set-up. It would be interesting to compute in this last case other canonical commutators, involving canonical momenta, and study similar exchanges for those.

The entanglement between coordinates and duals in the last background is really typical of non-geometry, and was reinterpreted in different formulations such as the T-fold \cite{Hull:2004in}, the doubled geometry \cite{Dabholkar:2005ve}, and also the generalized complex geometry where one would argue for an entanglement of vectors and one-forms \cite{Grana:2008yw}. A way to rephrase this idea is by saying that the toroidal fibre is fuzzy in the non-geometric case, and it becomes impossible to determine precisely the string position in these directions. Actually, the (point particle) position concept may not even make sense for a non-geometric background, but the string still knows how to behave in such a space, thanks to its extension and additional symmetries. The indetermination of its position in the non-geometric case gives a quantum mechanical uncertainty relation
\begin{equation}
(\Delta Z^1)^2(\Delta Z^2)^2\geq H^2\, \, \langle N^3\rangle^2\, ,\label{eq:uncertainty}
\end{equation}
which is in agreement with the non-vanishing commutator \eqref{eq:resultccl}.

\interfootnotelinepenalty=10000

The difference with standard quantum mechanics is that the uncertainty here is a static phenomenon, no momentum is involved but rather a second coordinate. Indeed, a closed string winding the base circle would probe the (fuzzy) fibre with its non-trivial monodromy due to the non-geometry, without even moving. This is in contrast to a point particle (or a non-wound closed string), having still a well-defined position in the base, which therefore can only test the local geometry of the fibre, and does not perceive the global aspects responsible for the non-geometry and non-commutativity. This explains the presence of the winding (or dual momentum) in our result \eqref{eq:resultccl} and \eqref{eq:uncertainty}. This discussion and the above relation provide an interesting, though heuristic, origin to the non-commutativity observed.\footnote{Note that another stringy feature than the winding has been pointed out to explain the non-commutativity: its origin was traced back in section \ref{sec:origin} to some 
oscillator terms in the string expansion, meaning that only these stringy modes are responsible for the non-vanishing right-hand side of \eqref{eq:resultccl}.}\\

\interfootnotelinepenalty=100

Finally, let us discuss the last step of the T-duality chain \eqref{eq:TdualityChain}, involving an $R$-flux, for which non-associativity has been argued to arise as mentioned in the introduction. In addition to the non-commutativity relation \eqref{commcloseda}, we also conjectured in \cite{Andriot:2012an}, following \cite{Blumenhagen:2010hj,Lust:2010iy,Blumenhagen:2011ph}, that the non-associativity of the $R$-flux background should be given by the associator
\beq
[{\cal X}^3, [{\cal X}^1, {\cal X}^2]]_{{\rm closed}} + \mbox{perm} \sim R^{123} \ . \label{eq:assoc}
\eeq
As discussed for the $Q$-flux in section \ref{sec:Qeffact}, to make sense of this relation one needs an expression for the flux in terms of the background fields. This was discussed for the $R$-flux in \cite{Andriot:2012wx,Andriot:2012an} together with an effective action for such a background.\footnote{To derive the $R$-flux effective action, we used the field redefinition \eqref{fieldredef}. In  \cite{Blumenhagen} a different field redefinition leading to another effective action for the non-geometric $R$-flux was proposed.} So let us first recall the procedure to get the $R$-flux, before coming back to the associator.

This last non-geometric situation is obtained after performing a T-duality in the direction $Z^3$ of the $Q$-flux background. Since the target space fields depend on $Z^3$, the Buscher rules do not apply anymore. However, as discussed in \cite{Andriot:2012an}, double field theory \cite{DFT} has a proposal on how to T-dualise along a direction which is not an isometry: we just need here to formally replace the coordinate $Z^3$ by its dual coordinate $\hat{Z}^3$, in analogy to the replacement of the momentum $p^3$ by its dual quantity the winding $N^3$. Performing this replacement in \eqref{eq:tfields}, we simply obtain for $\beta_R$
\beq \label{eq:tfields1}
\beta_R =  \begin{pmatrix}  0 & H \hat{Z}^3 & 0 \\ -H \hat{Z}^3 & 0 & 0 \\ 0 & 0 & 0 \end{pmatrix}
\ .
\eeq
The $R$-flux proposed \cite{Andriot:2012wx,Andriot:2012an} has the general form $R^{\mu\nu\rho}  =  3\tilde{D}^{[\mu}\beta^{\nu \rho]} = 3\big(\tilde{\partial}^{[\mu}\beta^{\nu\rho]}+\beta^{\varsigma[\mu}\partial_{\varsigma}\beta^{\nu\rho]}\big) $ (see also \cite{Dall'Agata:2007sr, Aldazabal:2011nj}), where $\tilde{\del}$ denotes the derivative with respect to the dual coordinate. It is straightforward to see that the second term does not contribute, while the first gives
\beq\label{eq:RH}
R^{123}  = H  \; .
\eeq

Let us now come back to the associator \eqref{eq:assoc}. Before considering the last background, let us start by saying a word on the $Q$-flux situation. For the latter, the first term on the LHS of \eqref{eq:assoc} can be computed to linear order in $H$: indeed, using \eqref{X0X0}, we get in the limit \eqref{eq:resultccl}
\beq
[Z^3, [Z^1, Z^2]] = [Z_0^3, [Z^1, Z^2]|_H ] + [ H Z_H^3, [Z_0^1, Z_0^2]] =  - \frac{\i}{2} \frac{\pi^2}{3} H [Z_0^3 , N^3]  = 0 \ .\nn
\eeq
In addition for this background, it is simple to see from the fields \eqref{eq:tfields} and the definition of the $R$-flux that $R^{123}$ is zero. Therefore, we are rather close to satisfy the associator condition \eqref{eq:assoc}. However, the permutation terms involve $H$-order commutators between $(1,3)$ and $(2,3)$ coordinates and associated modes. We have not determined these commutators here, and it is not clear if they could all be set to zero, in view of the non-trivial commutators \eqref{dtYdtYH31} and \eqref{dtYdtYH32}. Even if we do expect that the permutation terms vanish in the limit, we leave the full computation for future work.

The more interesting case for this associator is the $R$-flux background, with non-zero flux \eqref{eq:RH}. As above, we can say something on the first term of the associator but not on the others. The reason is this time deeper: the T-duality relations among the coordinates \eqref{Tdcoordrel} were derived using the Buscher procedure, which needs an isometry. We therefore have here no guideline to determine what is the mode expansion of the dual $\hat{Z}^3$, implying in particular that we cannot compute its commutators, such as those of the permutation terms. This problem can be solved at zeroth order in $H$: there, we only have the free string and it is therefore natural to take $\hat{Z}_0^3 \equiv \tilde{Z}_0^3$, as defined in \eqref{eq:sol0LR}. To avoid any confusion, let us denote the ``coordinates'' $W^{\mu}$ for the $R$-flux background, meaning after the (formal) T-duality in the third direction. We propose to define these coordinates as follows: $W^{1,2}(p^3_{W}) = Z^{1,2}(N^3)$, $W^3_0=\tilde{Z}_0^3$,
 where $p^3_W$ is now the zeroth order momentum for $W^3_0$, and $W^{1,2}(p^3_W)$ means that these coordinates now depend on $p^3_W$ instead of the winding (dual momentum) $N^3$ (i.e. we replace one by the other and the closed string boundary conditions of the $R$-flux background are now determined by $p^3_W$).
This allows us to compute the first term in the limit \eqref{eq:resultccl}
\beq
[W^3, [W^1, W^2]] = [W_0^3, [W^1, W^2]|_H ] + [H W_H^3, [W^1_0, W^2_0]]  =  - \frac{\i}{2} \frac{\pi^2}{3} H [W_0^3 , p^3_W]  = \frac{1}{4} \frac{\pi^2}{3} H \ ,\nn
\eeq
where we used \eqref{xpLR}. In view of the associator \eqref{eq:assoc}, it is nice to get a non-zero result, proportional to $H$, now understood as the $R$-flux. We leave a full study for the future.

An $R$-flux background is usually thought of as being not even locally geometric \cite{Dabholkar:2005ve}. For such a situation, one could then indeed consider that the closed string ``coordinates'', which should be really viewed as fields living on the two-dimensional string world-sheet, become non-associative. The precise mathematical structure of these non-associative spaces is very interesting but still largely unknown, and hence deserves more investigations in the future.

\subsection*{Acknowledgements}

We would like to thank B. Basso, A. Chatzistavrakidis, A. Deser, F. Rennecke, and in particular I. Bakas, for useful discussions at various stages of this project. The research of ML is supported by the Swedish Research Council (VR) under the contract 623-2011-7205.

\newpage

\begin{appendix}

\section{Rescaling and summary of notations}\label{ap:not}

For the three backgrounds we consider, the target space metric $G_{\mu\nu}$ and the $B$-field $B_{\mu\nu}$ depend on the three radii $R_{\mu=1,2,3}$. As a consequence, the world-sheet equations of motion and the string coordinates classical solutions would also depend on them. To simplify formulas, we perform the rescaling given in table \ref{tab:resc} for each of the backgrounds. It is defined on any object $V$ with a (curved space) index $\mu$, so in particular on the metric and $B$-field, on the $H$-flux component, on the string coordinates, and on the winding numbers.
{\renewcommand{\arraystretch}{1.5}
\begin{table}[H]
\begin{center}
\begin{tabular}{|c||c|}
\hline
Background & Rescaling \\
\hline
\hline
Torus + $H$-flux & $V^{\mu} \rightarrow \frac{1}{R_{\mu}} V^{\mu}\ , \ V_{\mu} \rightarrow R_{\mu} V_{\mu}$ (no sum) \\
\hline
Tw. torus & $V^{1} \rightarrow R_{1} V^{1}\ , \ V_{1} \rightarrow \frac{1}{R_{1}} V_{1}$, and for $\mu=2,3$ as for the torus \\
\hline
Non-geom. & $V^{1,2} \rightarrow R_{1,2} V^{1,2}\ , \ V_{1,2} \rightarrow \frac{1}{R_{1,2}} V_{1,2}$, and for $\mu=3$ as for the torus\\
\hline
\end{tabular}
\caption{Rescalings.}\label{tab:resc}
\end{center}
\end{table}}
\vspace{-0.2in}
Let us detail the rescaling of the $H$-flux component (first background). In curved space, one has $H_{123}=H$, which therefore gives $H\rightarrow H R_1 R_2 R_3$, or in other words, $H_{{\textrm old}} = H_{{\textrm new}} R_1 R_2 R_3$. This has the important consequence of simplifying the assumption \eqref{eq:approx}, so that after the rescaling, all fields are expanded in $H$ only.

Applying these rules to the target space fields of the different backgrounds, namely \eqref{Ametric} and \eqref{Abfield}, \eqref{metric}, and \eqref{eq:nongeofields}, has the simple effect of erasing all radii. The same is true for the world-sheet equations of motion. Note also that all contractions invariant, in particular differential forms, or the squared line element $\d s^2 = G_{\mu \nu} \d {\cal X}^{\mu} \d {\cal X}^{\nu}$. Finally, the T-duality relations \eqref{Tdcoordrel}, the string coordinates boundary conditions, and the canonical commutation relations, are as well invariant under this rescaling. From section \ref{sec:Tdual} on, we only use rescaled quantities. The table \ref{tab:rescandH} summarises the rescaled and approximated target space fields.\\

We denote the generic string coordinate by ${\cal X}^{\mu}$, and introduce different notations for each of the three T-dual backgrounds, namely $X^{\mu}$, $Y^{\mu}$, and $Z^{\mu}$ respectively. The index $\mu$ denotes the three dimensions $1,2,3$ of the target space that we consider; the third one is special and always corresponds to the base circle of a toroidal fibration. As all the fields, the string coordinate is expanded (after rescaling) up to second order terms in $H$, so we write it as
\beq
{\cal X}^{\mu}(\tau, \sigma) = {\cal X}_0^{\mu}(\tau, \sigma) + H{\cal X}_H^{\mu}(\tau, \sigma) + {\cal O}(H^2)\ .
\eeq
The zeroth and first order contributions have a classical mode expansion, once the equations of motion with boundary conditions are solved. The modes of $Y^{\mu}$, i.e. for the twisted torus, are those which appear mostly in the paper. The zeroth order $Y_0^{\mu}$ turns out to be a free string, so it depends on the standard constants: the center of mass position $y^{\mu}$, the momentum $p^{\mu}$, the winding $N^{\mu}$, and the oscillators $\widetilde{\alpha}_n^{\mu}, \alpha_n^{\mu}$. The $H$-order piece $Y_H^{\mu}$ has additional constants playing a similar role: the $H$-order $y_H^{\mu}$, $p_H^{\mu}$, and the oscillators $\tg_n^{\mu}, \g_n^{\mu}$. The expression for $Z^{\mu}$ is built from the one of $Y^{\mu}$; the only new constants appearing there are the center of mass positions $z^{\mu}, z_H^{\mu}$. The solution for $X^{\mu}$ is computed from scratch, and we obtain similar constants, denoted in the same manner but with an index ${}_X$, except for $x^{\mu}, x_H^{\mu}$. T-duality then fixes the modes of
$X^{\mu}$ in terms of those of $Y^{\mu}$ (see section \ref{sec:solutions}), so we rather express things in terms of the latter.

\section{Target space T-duality}\label{ap:Td}

One convenient way to perform a T-duality along a direction $\iota$ on the target space fields goes as follows. One should first consider the generalized metric $\hhh$
\beq
\hhh=\left(\begin{array}{c|c} G-BG^{-1}B & BG^{-1} \\ \hline -G^{-1}B & G^{-1} \end{array}\right) \ . \label{eq:genmet}
\eeq
The T-dual generalized metric $\hat{\hhh}$ (where one can read form its entries the T-dual metric and $B$-field) is then obtained, thanks to the action of the matrix $T$, as
\beq
\hat{\hhh}=T \hhh T \ , \ \mbox{where}\ T = \left(\begin{array}{c|c} 1-m_{\iota} & m_{\iota} \\ \hline m_{\iota} & 1-m_{\iota}\\ \end{array}\right) \ , \ \mbox{with}\ m_{\iota} = {\tiny \left(\begin{array}{c|ccc} 1 & 0 & \dots & 0  \\ \hline 0 & 0 & \dots & 0 \\ \vdots & \vdots & \ddots & \vdots \\ 0 & 0 & \dots & 0 \end{array}\right)} \ , \label{BuscherH}
\eeq
where the non-trivial entry of $m_{\iota}$ is along the direction $\iota$ (arbitrarily placed here in the top left corner). One can show that the expressions of the T-dual fields obtained either by the Buscher rules \eqref{BuscherE}, or this procedure \eqref{BuscherH}, precisely match, so the two procedures are equivalent.

To illustrate this method, let us focus on the two-torus with $B$-field as given in \eqref{eq:taurhotorus} (we do not need to consider here the third direction ${\cal X}^3$, which remains invariant under the following transformations). We can rewrite the torus fields in terms of matrices as
\beq
G =  \frac{\mbox{Im} \rho}{\mbox{Im} \tau} \begin{pmatrix} 1 & \re \tau \\ \re \tau & |\tau|^2 \end{pmatrix}
\ ,\ \ B= \begin{pmatrix} 0 & -\re \rho \\ \re \rho & 0 \end{pmatrix} \ . \label{metB}
\eeq
We can then compute the T-duals of this configuration, by first constructing the generalized metric $\hhh$ defined in \eqref{eq:genmet}
\beq
\hhh= \frac{1}{\im \rho \im \tau}
\left(\begin{array}{cc|cc}
|\rho|^2 &  |\rho|^2 \re \tau & \re \rho \re \tau & -\re \rho \\
|\rho|^2 \re \tau & |\rho \tau|^2 & \re \rho |\tau|^2 & -\re \rho \re \tau \\
\hline
\re \rho \re \tau & \re \rho |\tau|^2 & |\tau|^2 & -\re \tau \\
-\re \rho & -\re \rho \re \tau & -\re \tau & 1
\end{array}\right) \ .
\eeq
The T-duals are obtained by applying the proper T-duality operators as in \eqref{BuscherH}. Namely, we obtain the following T-dual fields
\bea
\mbox{T-duality along ${\cal X}^1$ :} & \quad \ G=\frac{\im \tau}{\im \rho}  \begin{pmatrix} 1 & \re \rho \\ \re \rho & |\rho|^2  \end{pmatrix} \ , \ B=\begin{pmatrix} 0 & -\re \tau \\ \re \tau & 0 \end{pmatrix} \nn\\
& \qquad \qquad \qquad \qquad \qquad \qquad \qquad \qquad \qquad \qquad  \Leftrightarrow \tau \leftrightarrow \rho \label{apTx}\\
\mbox{T-duality along ${\cal X}^2$ :} & \quad \ G=\frac{\im \tau |\rho|^2}{|\tau|^2 \im \rho} \begin{pmatrix} 1 & -\frac{\re \rho}{|\rho|^2} \\ -\frac{\re \rho}{|\rho|^2} & \frac{1}{|\rho|^2}  \end{pmatrix} \ , \ B=\frac{\re \tau}{|\tau|^2} \begin{pmatrix} 0 & 1 \\ -1 & 0 \end{pmatrix} \nn\\
& \qquad \qquad \qquad \qquad \qquad \qquad \qquad  \Leftrightarrow \tau \to -\frac{1}{\rho} \ ,\ \rho \to -\frac{1}{\tau} \label{apTy} \\
\mbox{T-duality along ${\cal X}^1$ and ${\cal X}^2$ :} & \quad \ G=\frac{\im \rho |\tau|^2}{|\rho|^2 \im \tau} \begin{pmatrix} 1 & -\frac{\re \tau}{|\tau|^2} \\ -\frac{\re \tau}{|\tau|^2} & \frac{1}{|\tau|^2}  \end{pmatrix} \ , \ B=\frac{\re \rho}{|\rho|^2} \begin{pmatrix} 0 & 1 \\ -1 & 0 \end{pmatrix} \nn\\
& \qquad \qquad \qquad \qquad \qquad \qquad \qquad  \Leftrightarrow \tau \to -\frac{1}{\tau}\ , \ \rho \to -\frac{1}{\rho} \ ,\label{apTxTy}
\eea
where we read the corresponding exchanges of $\tau$ and $\rho$ by comparing with \eqref{metB}. We can then easily use these formulas to derive the T-dual background fields we consider, namely \eqref{Ametric} and \eqref{Abfield}, \eqref{metric}, and \eqref{eq:nongeofields}, or equivalently \eqref{rhotautorusH}, \eqref{rhotautwtorus}, and \eqref{rhotaunongeo}.

\section{Global properties and boundary conditions in terms of monodromies}\label{sec:monod}

In this appendix, we study the global properties of the three T-dual backgrounds presented in section \ref{sec:classstring}, and express them in terms of monodromies (see \cite{Dall'Agata:2007sr} for a related discussion). We then make use of this formulation, together with a doubled formalism, to discuss the boundary conditions of the string coordinates.

\subsection{Geometry and non-geometry from monodromies}

The three backgrounds considered in section \ref{sec:classstring} can be described in a convenient uniform manner: they all take the form, at least locally, of the fibration of the torus $T^2$ with (real) coordinates ${\cal X}^{1,2}$ over the circle $S^1$ in the ${\cal X}^3$-direction. A two-torus with $B$-field is parametrized by two complex parameters known as the complex structure $\tau={\frac{G_{12}}{G_{11}}}+\i\ {\frac{V}{G_{11}}}$ and the complexified K\"ahler class $\rho=-B_{12}+\i\ V$, where $V$ denotes the volume of the two-torus.\footnote{ $V$ is the square root of the absolute value of the determinant of the $T^2$ metric, i.e. in our notations {\small $V=\sqrt{|G_{11} G_{22}-(G_{12})^2|}$}.} Therefore, our three backgrounds can be reexpressed as
\beq
ds^2 = \frac{\im \rho}{\im \tau} \left|d{\cal X}^1 + \tau d{\cal X}^2\right|^2 + R_3^2 \left(\d{\cal X}^3 \right)^2 \ ,\ B_{12}=-\re \rho \ ,\label{eq:taurhotorus}
\eeq
with
\bea
\mbox{Torus with $H$-flux:} &\quad \ \tau= \i\ \frac{R_2}{R_1}\ , \  \rho=-H {\cal X}^3 + \i\ R_1 R_2 \label{rhotautorusH} \\
\mbox{Twisted torus:} &\quad \ \tau=-H {\cal X}^3 + \i\ R_1 R_2 \ , \  \rho=\i\ \frac{R_2}{R_1} \label{rhotautwtorus} \\
\mbox{Non-geometric background:} &\quad \ \tau= \i\ \frac{R_1}{R_2}  \ , \  \rho=\frac{1}{H {\cal X}^3 - \i\ R_1 R_2} \ , \label{rhotaunongeo}
\eea
where the fibration can be seen via the dependence of $\tau$ and $\rho$ on the base coordinate ${\cal X}^3$.

The global structure of the fibrations can be characterised by the monodromy conditions, which indicate how the $T^{2}$ is glued together when moving around the base $S^1$:
\begin{equation}\label{m1}
{\cal X}^3\rightarrow {\cal X}^3 +2\pi \quad \Rightarrow \quad T^2({\cal X}^3)\rightarrow T^2({\cal X}^3+2\pi)\, .
\end{equation}
In order to obtain a consistent string background, the gluing conditions of $T^2$ have to be part of the string symmetry group. The symmetries of a string on a two-torus, parametrized as above, are well-known, and are essentially captured by two $SL(2,\mathbb{Z})$ factors, acting on $\tau$ and $\rho$ separately, as
\beq
\tau  \to \frac{a \tau + b}{c  \tau + d} \ , \ A=\begin{pmatrix} a & b\\ c & d\\ \end{pmatrix} \in SL(2,\mathbb{Z})_{\tau} \ ,\ \rho  \to \frac{a' \rho + b'}{c'  \rho + d'} \ , \ A'=\begin{pmatrix} a' & b'\\ c' & d'\\ \end{pmatrix} \in SL(2,\mathbb{Z})_{\rho} \ .\label{eq:tautransfo}
\eeq
Therefore, consistent string backgrounds are obtained from the configuration \eqref{eq:taurhotorus}, if the monodromy relation \eqref{m1} can be written as
\beq\label{m2}
{\cal X}^3\rightarrow {\cal X}^3 +2\pi \quad \Rightarrow \quad \tau({\cal X}^3 +2\pi)=\frac{a~\tau({\cal X}^3)+b}{c~\tau({\cal X}^3)+d} \ ,\ \rho({\cal X}^3 +2\pi)=\frac{a'~\rho({\cal X}^3)+b'}{c'~\rho({\cal X}^3)+d'} \ ,
\eeq
in terms of two $SL(2,{\mathbb Z})$ transformations.

This reasoning however does not distinguish geometric or non-geometric backgrounds; in particular the latter could occur since some T-dualities are part of this string symmetry group, as can be seen in appendix \ref{ap:Td}. So let us say a word on this distinction. First, monodromies given in terms of $SL(2,{\mathbb Z})_\tau$, i.e. the transformations on the complex structure $\tau$, are always geometric. Indeed, one can verify that the transformation $A$ in \eqref{eq:tautransfo} can be reproduced precisely by the global diffeomorphism $\tilde{A}$ on $T^2$, given by
\beq
\begin{pmatrix}
\d {\cal X}^1\\
\d {\cal X}^2\\
\end{pmatrix}
\to
\begin{pmatrix}
d & b\\
c & a\\
\end{pmatrix} \begin{pmatrix}
\d {\cal X}^1\\
\d {\cal X}^2\\
\end{pmatrix}
\ ,\ \tilde{A}= \begin{pmatrix}
d & b\\
c & a\\
\end{pmatrix} \ , \label{eq:SL2coord}
\eeq
i.e. the metric \eqref{eq:taurhotorus} transforms in the same way by either transformation. In other words, any $SL(2,{\mathbb Z})_\tau$ monodromy $A$ can be compensated by a discrete identification of the coordinates, given by the inverse $\tilde{A}^{-1}$ of \eqref{eq:SL2coord}; the latter also leaves the $B$-field invariant. This way, the gluing of the $T^2$ is fine, and such backgrounds are geometric. One can verify from \eqref{rhotautwtorus} that it is precisely what happens for the twisted torus, as indicated by the identifications \eqref{monod1}.

On the contrary, non-geometric backgrounds in general correspond to those monodromy transformations which act non-trivially as $SL(2,{\mathbb Z})_\rho$ on the K\"ahler parameter $\rho$. To see that more precisely, let us first make the following distinction among the possible monodromies $A'$ of \eqref{eq:tautransfo}. We define the order $n \in \mathbb{N}^*$ of $A'$ as the minimal value allowing for $A'^{\ n} = \id$. Then, we distinguish
\begin{itemize}
\item{Constant monodromy: $n=1$, i.e. $\rho({\cal X}^3 + 2 \pi) =\rho({\cal X}^3)$}
\item{Elliptic monodromy: $n>1$ but finite; for instance the elliptic inversion with $n=4$: $\rho({\cal X}^3 + 2 \pi) = \frac{-1}{\rho({\cal X}^3)}$}
\item{Parabolic monodromy: a finite order does not exist; for instance the constant shift $\rho({\cal X}^3 + 2 \pi) =\rho({\cal X}^3) + b'$, $b' \in \mathbb{Z}^*$.}
\end{itemize}
As a side remark, note that the last two examples actually generate the whole $SL(2,{\mathbb Z})$ group.

Constant monodromies obviously leave us with geometric backgrounds; some parabolic also do. For instance the constant shift is only a gauge transformation on the $B$-field. An example of it is given by the torus with $H$-flux, as can be seen from \eqref{rhotautorusH}. Other parabolic monodromies can give non-geometric backgrounds. For example, one obtains from \eqref{rhotaunongeo}
\beq
\rho({\cal X}^3+2\pi)= \frac{\rho({\cal X}^3)}{1+2\pi H \rho({\cal X}^3)} \ ,
\eeq
which is a parabolic monodromy: it has the form of the ``transpose'' of a constant shift. Finally, the elliptic monodromy rather gives a non-geometric background, because the resulting change on the physical fields is really non-trivial. For instance, the example given above can correspond to a T-duality, as in \eqref{apTxTy}.

\subsection{Boundary conditions of (doubled) string coordinates}

The monodromy transformations on the background parameters can be translated into the closed string boundary conditions, that we use when solving the equations of motion. Namely, going around the closed string in the ${\cal X}^3$-direction by doing $\sigma \to \sigma + 2 \pi$, i.e. considering its winding (given by $N^3$)
\begin{equation}
{\cal X}^3(\tau,\sigma)\rightarrow {\cal X}^3(\tau,\sigma+2\pi) = {\cal X}^3(\tau,\sigma) + 2\pi N^3\, ,
\end{equation}
corresponds to having $N^3$ times the transformation \eqref{m2}. So it precisely induces monodromy transformations of the form described above
\beq
\tau({\cal X}^3(\tau,\sigma+2\pi))=\frac{a~\tau({\cal X}^3(\tau,\sigma))+b}{c~\tau({\cal X}^3(\tau,\sigma))+d} \ , \ \rho({\cal X}^3(\tau,\sigma+2\pi))=\frac{a'~\rho({\cal X}^3(\tau,\sigma))+b'}{c'~\rho({\cal X}^3(\tau,\sigma))+d'} \ .\label{m3}
\eeq
From those we would like to deduce how the coordinates ${\cal X}^{1,2}$ of the two-dimensional torus transform, and understand it as the closed string boundary conditions in these directions. To deduce the transformation of ${\cal X}^{1,2}$ from a $\tau$-monodromy given in \eqref{m3}, we can proceed as discussed around \eqref{eq:SL2coord}. Thus, for a transformation $A$ of $\tau$ given in \eqref{eq:tautransfo} and \eqref{m3}, the linear action of $\tilde{A}^{-1}$ on ${\cal X}^{1,2}$ is the one giving their boundary conditions.

We want to embed this idea in a doubled formalism inspired by the T-fold description \cite{Hull:2004in}. To that end, we double the two toroidal coordinates and introduce the vector $\vec{\cal X}=({\cal X}^1,{\cal X}^2,\hat {\cal X}^1,\hat {\cal X}^2)$, where $\hat{\cal X}^{\mu}$ are the dual coordinates (denoted as in \eqref{Tdcoordrel}). The transformation of $\vec{\cal X}$ is now given by the linear action of an $O(2,2,{\mathbb Z})$ element $g$, and we consider the four-dimensional representation of $g^T J g =J$, where $J$ is given by two off-diagonal $\id_2$. So the closed string boundary conditions are now written as
\beq
{\cal X}^3(\tau,\sigma)\rightarrow {\cal X}^3(\tau,\sigma+2\pi) \quad \Rightarrow \quad \vec{\cal X}(\tau,\sigma) \rightarrow \vec{\cal X}(\tau,\sigma+2\pi)= g~\vec{\cal X}(\tau,\sigma)\ , \label{m4}
\eeq
where the $O(2,2,{\mathbb Z})$ matrix $g$ is determined by the monodromy transformations \eqref{m3} (in other words, $g$ is given by an embedding of $SL(2,{\mathbb Z})_\tau\times SL(2,{\mathbb Z})_\rho$ in $O(2,2,{\mathbb Z})$). For instance, from what we explained above, a $\tau$-monodromy given by a matrix $A$ fixes
\beq
g_{\tau}=\left(\begin{array}{cc} \tilde{A}^{-1} & 0_2 \\  0_2 & \tilde{A}^{T} \end{array}\right) \ , \label{eq:gtau}
\eeq
where the completion with $\tilde{A}^{T}$ is given by the $O(2,2,{\mathbb Z})$ condition.

Boundary conditions due to $\rho$-monodromies are determined differently than the $\tau$ ones. Consider the complexified momentum vector {\small $\Uppi=\Uppi^1+\i\  \Uppi^2=\frac{1}{\sqrt{\im\rho \im\tau}}(p^2+\bar\tau p^1+\bar\rho(N^1-\bar\tau N^2))$}, and find the transformation of the momentum and winding $p^{\mu}, N^{\nu}$, which compensates the one on $\rho$, so that {\small $|\Uppi|$} remains invariant. This transformation on $(p^1, p^2, N^1, N^2)$ is then identified with $g^T$. Let us give two examples
\beq
\rho\rightarrow\rho+b':\ g_{b'}=
\begin{pmatrix}1&0&0&0\\ 0&1&0&0\\0&-b'&1&0\\b'&0&0&1 \end{pmatrix}\ ,
\qquad \rho\rightarrow-1/\rho:\ g_i=
\begin{pmatrix}0&0&0&-1\\ 0&0&1&0\\0&-1&0&0\\1&0&0&0
\end{pmatrix}\ .\label{eq:grho}
\eeq
The constant shift compensated by $g_{b'}^T$ actually leaves the whole vector {\small $\Uppi$} invariant, while the elliptic inversion compensated by $g_i^T$ only leaves {\small $|\Uppi|$} invariant. Considering $\i g_i$ would equally well leave {\small $|\Uppi|$} invariant; however such an action on the coordinates is not clear: this may indicate that an elliptic inversion of $\rho$ alone should not be considered. We will nevertheless make use of $\i g_i$ in what follows when composing several inversions.

Now we are eventually ready to explicitly relate the monodromy conditions and the closed string boundary conditions for the three backgrounds considered in this paper.

\subsubsection*{Closed string boundary conditions for the torus with $H$-flux}

Here the monodromy corresponds to a constant shift in the $B$-field: we read from \eqref{rhotautorusH} that $\rho({X}^3(\tau,\sigma+2\pi))=\rho({X}^3(\tau,\sigma))-2\pi H N^3$. Therefore, using \eqref{eq:grho}, the matrix $g$ is given by $g_{-2\pi H N^3}$, leading to the following closed string boundary conditions
\bea
X^{1}(\tau,\sigma+2\pi) &= X^{1}(\tau,\sigma)  \; ,\nonumber\\
X^{2}(\tau,\sigma+2\pi) &= X^{2}(\tau,\sigma)  \; ,\nonumber\\
\hat X^{1}(\tau,\sigma+2\pi) &= \hat X^{1}(\tau,\sigma) + 2\pi HN^3X^{2}(\tau,\sigma)\; ,\nonumber\\
\hat X^{2}(\tau,\sigma+2\pi) &= \hat X^{2}(\tau,\sigma) -2\pi HN^3X^{1}(\tau,\sigma)\;  ,\nonumber\\
X^{3}(\tau,\sigma+2\pi)& = X^{3}(\tau,\sigma) + 2 \pi N^3 \; .\label{eq:bdyXmonod}
\eea
For $X^1,\ X^2$, this is in agreement with \eqref{eq:bdyX}, up to constant shifts such as winding, a point we will come back to. In addition we find that the dual coordinates transform non-trivially.

\subsubsection*{Closed string boundary conditions for the twisted torus}

Here the monodromy corresponds to a constant shift of the complex structure: we read from \eqref{rhotautwtorus} that $\tau({Y}^3(\tau,\sigma+2\pi))=\tau({Y}^3(\tau,\sigma))-2\pi H N^3$. The monodromy transformation $A$ is then given as in \eqref{eq:tautransfo}, and the corresponding $\tilde{A}$ is given as in \eqref{eq:SL2coord}. We deduce from the latter the matrix $g$ given in \eqref{eq:gtau}, leading to the following closed string boundary conditions
\bea
Y^1(\tau,\sigma+2\pi) &= Y^1(\tau,\sigma)  + 2\pi H N^3 Y^2 (\tau,\sigma) \, ,\nonumber\\
Y^2(\tau,\sigma+2\pi) &= Y^2(\tau,\sigma) \, , \nonumber\\
\hat Y^{1}(\tau,\sigma+2\pi) &= \hat Y^{1}(\tau,\sigma)\, ,\nonumber\\
\hat Y^{2}(\tau,\sigma+2\pi) &= \hat Y^{2}(\tau,\sigma) -2\pi H N^3 \hat Y^{1}(\tau,\sigma)\; ,\nonumber\\
Y^{3}(\tau,\sigma+2\pi)& = Y^{3}(\tau,\sigma) + 2 \pi N^3 \; .\label{eq:bdyYmonod}
\eea
This is again in agreement with \eqref{bdy1} and \eqref{bdy2}, up to constant shifts such as winding.

\subsubsection*{Closed string boundary conditions for the non-geometric background}

As already discussed, we have here a parabolic monodromy in $\rho$: we read from \eqref{rhotaunongeo} that
\begin{equation}
\rho({Z}^3(\tau,\sigma+2\pi))=\frac{\rho({Z}^3(\tau,\sigma))}{1+2\pi HN^3~\rho({Z}^3(\tau,\sigma))}\, .
\end{equation}
To obtain the corresponding matrix $g$, one can notice that this monodromy is actually the composition of an elliptic inversion, a constant shift by $-2\pi HN^3$, and again an elliptic inversion. Therefore, the $g$ to consider here is given by $g=\i g_i\ g_{-2\pi HN^3}\ \i g_i$, as one can obtain using \eqref{eq:grho} and the discussion around there. The result is the transpose of $g_{2\pi HN^3}$; it is therefore very close to the one used for the torus with $H$-flux. This matrix finally leads to
\bea
Z^1(\tau,\sigma+2\pi) &= Z^1(\tau,\sigma)  + 2\pi H N^3 \hat Z^2 (\tau,\sigma) \, ,\nonumber\\
Z^2(\tau,\sigma+2\pi) &= Z^2(\tau,\sigma)  - 2\pi H N^3 \hat Z^1 (\tau,\sigma)\, , \nonumber\\
\hat Z^{1}(\tau,\sigma+2\pi) &= \hat Z^{1}(\tau,\sigma)\, ,\nonumber\\
\hat Z^{2}(\tau,\sigma+2\pi) &= \hat Z^{2}(\tau,\sigma) \;  ,\nonumber\\
Z^{3}(\tau,\sigma+2\pi)& = Z^{3}(\tau,\sigma) + 2 \pi N^3 \; .\label{eq:bdyZ monod}
\eea
This transformation is new, and should be understood, as before, up to constant shifts of the coordinates. It is consistent with the classical expressions of $Z^{\mu}$ that are derived using the T-duality relations among coordinates. The non-trivial match is in particular with the $Z^2$ boundary conditions given in \eqref{zztrans}.

It is worth noticing that this background is the only one for which the boundary conditions of the standard coordinates mix with the dual coordinates. This entanglement is a sign of non-geometry, but it is also reminiscent of the mixing of Neumann and Dirichlet boundary conditions for the open string, leading in that context to non-commutativity. This analogy is announcing the result of this paper, as we find closed string non-commutativity for this background.

\subsubsection*{Final comments}

We used here an argumentation based on preserving either the target space fields or the above {\small $|\Uppi|$} when going through monodromies. An alternative derivation of the closed string boundary conditions can be made using the doubled formalism of \cite{Nibbelink:2012jb} and requiring to preserve an action. In both cases, we obtain the (same) expected boundary conditions, up to shifts by constants such as winding. Here, the reason is clear, as we only consider a linear action on the coordinates \eqref{m4}. The argument leading to this linear action was actually based on the analogy with the $\tau$-monodromies discussed around \eqref{eq:SL2coord}. However, there, it was rather $\d {\cal X}^{\mu}$ than ${\cal X}^{\mu}$ itself which was acted on linearly. Following this path then opens the door to possible constant shifts of the coordinates, even though nothing fixes them in this procedure.

\section{Commutators of position zero modes}\label{sec:fixcom}

As discussed below \eqref{deldelmagic}, the classical integration constants $z^{1,2}$ and $z_H^{1,2}$ for the coordinates $Z^{1,2}$ lead, at the quantum level, to undetermined commutators present in the expression \eqref{eq:comfinalb} obtained for $[Z^1,Z^2]$. In this appendix, we provide arguments that fix those unknown commutators to specific values. We find these arguments reasonable and consistent with the study performed in this paper, although a priori other reasonings could as well be pursued.

To start with, one can note that the constants $z^{1,2}$ and $z_H^{1,2}$ are center of mass position coefficients at zeroth and first order. As such, they can be understood as the T-dual counterparts, along direction $2$, of respectively $y^{1,2}$ and $y_H^{1,2}$. Therefore, we propose that $z^1$, resp. $z_H^1$, has the same $H$-order commutators as $y^1$, resp. $y_H^1$; similarly, $z^2$, resp. $z_H^2$, has the same $H$-order commutators as $\tilde{y}^2$, resp. $\tilde{y}_H^2$. This result about the $H$-order commutators is rather close to the one we argued for at zeroth order (see above \eqref{eq:simplifZZ}, as well as \eqref{eq:Magda}). So we first obtain for the undetermined commutators
\bea
& \Big[z_H^1 , N^2 \Big] \equiv \Big[y_H^1 , N^2 \Big] = 0 \label{eq:N2z1H}\\
& \Big[p^2, z_H^1 \Big] + \Big[p_H^2, z^1 \Big] \equiv \Big[p^2, y_H^1 \Big] + \Big[p_H^2, y^1 \Big] = - \frac{\i}{2} y^3 \label{eq:p2z1H}\\
& \Big[z_H^1 , \alpha_{n\epsilon}^{2} \Big] + \Big[z^1,  \g_{n\epsilon}^{2} \Big] \equiv \Big[y_H^1 , \alpha_{n\epsilon}^{2} \Big] + \Big[y^1,  \g_{n\epsilon}^{2} \Big] = - \frac{1}{8n} \alpha^3_{n \epsilon} \ , \ \forall \epsilon,\ \forall n \neq 0 \ , \label{eq:a2z1H}
\eea
\bea
& \Big[z^1 , z_H^2\Big] + \Big[z_H^1 , z^2 \Big] \equiv \Big[y^1 , \tilde{y}_H^2\Big] + \Big[y_H^1 , \tilde{y}^2 \Big] \label{eq:yy}\\
& \Big[ p^1 , z_H^2\Big]  + \Big[p_H^1, z^2\Big] \equiv \Big[ p^1 ,\tilde{y}_H^2 \Big]  + \Big[p_H^1, \tilde{y}^2 \Big] \\
& \Big[ N^1 , z_H^2\Big] \equiv \Big[ N^1 , \tilde{y}_H^2 \Big]\\
& \Big[ \alpha_{n\epsilon}^{1} , z_H^2\Big] + \Big[ \g_{n\epsilon}^1  , z^2\Big] \equiv \Big[ \alpha_{n\epsilon}^{1} , \tilde{y}_H^2 \Big] + \Big[ \g_{n\epsilon}^1  , \tilde{y}^2 \Big]\ .\phantom{= - \frac{1}{8n} \alpha^3_{n \epsilon} \ , \ \forall \epsilon,\ \forall n \neq 0\ } \label{eq:super}
\eea
The value of the first three lines is simply determined, by using \eqref{eq:comzeromod}, \eqref{eq:comyD}, and \eqref{eq:comyCn}. On the contrary, the others remain undetermined. This is because the T-duality is here along ${\cal X}^2$: the information along direction $1$ is unchanged and can be used directly, while changes occur along direction $2$ (in particular here trading $y_{(H)}^2$ for $\tilde{y}_{(H)}^2$, leaving us with unknown commutators).

So we need information for the four commutators along this T-dualised direction. A first natural consideration is the left/right decomposition. The $H$-order constants always enter the homogeneous part of the expressions of the coordinates (see the first lines of \eqref{eq:Z1H} and \eqref{eq:tX2Hnew}, or of \eqref{finalsol}). Therefore, one can consider as for the zeroth order \eqref{eq:sol0LR} a left/right decomposition.\footnote{Such a decomposition was already mentioned for the first order in \eqref{solY0Y0} and \eqref{formZ2H}.} A simple assumption to be made is, as for the free string, that left/left and right/right commutators are equal while left/right vanish (see for instance \eqref{xpLR}). If we use this on the first undetermined commutator \eqref{eq:yy}, it simply vanishes. This is somehow expected: at zeroth order, center of mass positions commute, and it looks reasonable to get the same here, as in \eqref{eq:comzeromod}. We now turn to another undetermined commutator, \eqref{eq:super}. Using the
same decomposition within \eqref{eq:comyCn}, one fixes its value to \eqref{eq:a1z2H}. The remaining two unknown commutators can unfortunately not be determined using these arguments. One reason for that is the absence of winding at ${\cal O}(H)$ as discussed below \eqref{eq:fixpH1}, which makes the left/right decomposition of $p_H^1$ a bit ad-hoc. Therefore, we now present a second reasoning.\\

As the missing information is related to the T-duality along ${\cal X}^2$, we propose to compare  this situation with an analogous one, that is, the other T-duality considered in this paper. For the T-duality along ${\cal X}^1$, resp. along ${\cal X}^2$, one starts from $X^1$, resp. $Y^2$, to go to $Y^1$, resp. $Z^2$. The starting coordinates $X^1$ and $Y^2$ have similar boundary conditions given by a simple winding, while their T-dual counterparts $Y^1$ and $Z^2$ have more involved boundary conditions, as can be seen in \eqref{bdy1} and \eqref{zztrans}, or also \eqref{eq:bdyYmonod} and \eqref{eq:bdyZ monod}. The essential part of $Y^1$, resp. $Z^2$, boundary conditions, is a shift at $H$-order, by $Y_0^2$, resp. $-\tilde{Y}_0^1$. These shifts are also the essential part of each T-duality transformation: indeed, one can see from the T-duality relations between coordinates \eqref{eq:TdcoordXY} and \eqref{eq:TdcoordYZ} that the terms with explicit $H$ dependence correspond precisely to these same shifts. So
when comparing, at $H$-order, the two T-dualities performed in this paper, and looking at the various coordinates involved, we find an analogy between the two pairs $X^1_H, Y^2_0$ and $Y^2_H, -\tilde{Y}_0^1$. This is reminiscent of the idea that the order of the two T-dualities should not matter.

The relation between the two pairs is even more striking when looking at the explicit expressions for $X^1_H$ in \eqref{Xfinalsol} and $Y^2_H$ in \eqref{finalsol}. These two mode expansions are clearly mapped one to the other under the exchange of their $H$-order constants, together with the exchange of $Y^2_0$ and $-\tilde{Y}_0^1$ (or $\tilde{Y}^2_0$ and $-Y_0^1$) and equivalently of their zeroth order coefficients. This adds credit to the idea that the two pairs contain the same (physical) information. Therefore, we propose that ``physics remain unchanged'' under the map
\beq
Y^2_H \leftrightarrow X^1_H \ , \ Y^1_0 \leftrightarrow -\tilde{Y}^2_0 \ ,
\eeq
or equivalently the map of their modes, for instance $y^2_H \leftrightarrow x^1_H$. Let us extend the latter mildly towards $\tilde{y}^2_H \leftrightarrow \tilde{x}^1_H$. We now give meaning to the above statement by saying that a commutator preserves its value when going through this map. In other words, commutators involving the following modes keep the same value under the exchanges
\beq
\begin{array}{|lc|l}
 & & y^1 \leftrightarrow -\tilde{y}^2 \\
\tilde{y}^2_H \leftrightarrow y^1_H & & p^1 \leftrightarrow - N^2 \\
p^2_H \leftrightarrow p^1_{HX} & & N^1 \leftrightarrow - p^2 \\
\g^2_{m\epsilon} \leftrightarrow \g^1_{X m\epsilon} & & \alpha^1_{m\epsilon} \leftrightarrow - \epsilon\ \alpha^2_{m\epsilon} \ \forall \epsilon\ , \ \forall m \in \Z^* \ .
\end{array} \label{eq:map}
\eeq
We should have written $\tilde{y}^2_H \leftrightarrow \tilde{x}^1_H$ but we also trade the commutators involving $\tilde{x}^1_H$ for those involving $y^1_H$ (same argument as $z_H^2$ and $\tilde{y}^2_H$).

Let us now give an example: from this map, we find the following equality
\beq
[\tilde{y}^2, p^1_{HX}] + [\tilde{y}_H^2, N^1] = -[y^1, p^2_H] - [y_H^1, p^2] \ ,\label{eq:mapp1HX}
\eeq
where the RHS turns out to be known from \eqref{eq:comyD}. Using the latter together with \eqref{eq:fixpHX1}, the T-duality constraint fixing $p_{HX}^1$, one gets from \eqref{eq:mapp1HX}
\beq
[\tilde{y}_H^2, N^1]= \frac{\i \pi}{2} N^3 \ ,
\eeq
that gives a value to one of the undetermined commutators.

We turn to the other commutators. Using the T-duality constraints also fixes $p_H^1$ in \eqref{eq:fixpH1} and $\g^1_{Xm\epsilon}$ in \eqref{eq:fixg1Xm+} and \eqref{eq:fixg1Xm-}, and one gets
\beq
\Big[p_H^1, \tilde{y}^2 \Big] = 0 \ , \ \Big[ \g_{n\epsilon}^1  , \tilde{y}^2 \Big] = \epsilon \Big[ \g_{X n\epsilon}^1  , \tilde{y}^2 \Big] + \epsilon\ \frac{1}{4n} \alpha^3_{n \epsilon} \ ,
\eeq
that appear in two of the four unknown commutators we started with. Using these results, and then the map, one obtains the equalities
\bea
& \Big[ p^1 , \tilde{y}_H^2\Big]  + \Big[p_H^1, \tilde{y}^2\Big] = \Big[ y^1_H , N^2 \Big] \\
& \Big[ \alpha_{n\epsilon}^{1} , \tilde{y}_H^2\Big] + \Big[ \g_{n\epsilon}^1  , \tilde{y}^2\Big] = \epsilon \Big[ y_H^1, \alpha_{n\epsilon}^{2} \Big] + \epsilon \Big[ y^1, \g_{n\epsilon}^2 \Big] + \epsilon\ \frac{1}{4n} \alpha^3_{n \epsilon} \ , \ \forall \epsilon,\ \forall n \neq 0 \ .
\eea
Using \eqref{eq:comyCn} and \eqref{eq:comzeromod} finally gives values to the above, and so fixes two more commutators. Note that \eqref{eq:super} was already fixed by the left/right decomposition, and we recover here precisely the same value. Finally, the commutator \eqref{eq:yy} is not fixed by this map; it is simply mapped to itself. We need the above arguments to show that it vanishes.

Let us summarize the values obtained from these arguments for the four undetermined commutators:
\bea
& \Big[z^1 , z_H^2\Big] + \Big[z_H^1 , z^2 \Big] = 0 \label{eq:z1z2H+}\\
& \Big[ p^1 , z_H^2\Big]  + \Big[p_H^1, z^2\Big] = 0 \\
& \Big[ N^1 , z_H^2\Big] = - \frac{\i \pi}{2} N^3 \label{eq:N1z2H}\\
& \Big[ \alpha_{n\epsilon}^{1} , z_H^2\Big] + \Big[ \g_{n\epsilon}^1  , z^2\Big] = \epsilon\ \frac{1}{8n} \alpha^3_{n \epsilon} \ , \ \forall \epsilon,\ \forall n \neq 0 \ .\label{eq:a1z2H}
\eea
Finally, using these and \eqref{eq:N2z1H} - \eqref{eq:a2z1H} reduces the commutator of interest \eqref{eq:comfinalb} to \eqref{eq:resultsimpl}.

\end{appendix}

% ---------------------------------------- Bibliography

\newpage

\providecommand{\href}[2]{#2}\begingroup\raggedright\endgroup

\end{document}